\documentclass[useAMS,usenatbib]{mn2e}

\usepackage{graphicx}
\usepackage{amssymb}
\usepackage{amsmath}

\newcommand{\zebra}{\texttt{ZEBRA}}
\newcommand{\lzebra}{\texttt{\large{}ZEBRA}}

\newcommand{\apj}{\mbox{ApJ}}
\newcommand{\apjs}{\mbox{ApJS}}
\newcommand{\mnras}{\mbox{MNRAS}}
\newcommand{\aj}{\mbox{AJ}}
\newcommand{\aaa}{\mbox{AA}}

\newcommand{\LL}{\mathcal{L}}
\newcommand{\beq}{\begin{equation}}
\newcommand{\eeq}{\end{equation}}

\newcommand{\dd}{\mathrm{d}}

\newcommand{\pcm}{\textit{photometry-check mode}}
\newcommand{\pc}{\textit{photometry-check}}
\newcommand{\tom}{\textit{template-optimization mode}}
\newcommand{\teo}{\textit{template-optimization}}
\newcommand{\mlm}{\textit{Maximum-Likelihood mode}}
\newcommand{\bm}{\textit{Bayesian mode}}
\DeclareMathOperator{\Mag}{mag}

\title[\zebra{}: The Zurich Extragalactic Bayesian Redshift Analyzer]
 	{The Zurich Extragalactic Bayesian Redshift Analyzer   \\  
	({\sl ZEBRA}) and  its 1$^{st}$ application:  COSMOS }
\author[Feldmann et al.]{R. Feldmann$^{1}$, C.M. Carollo$^{1}$, C. Porciani$^{1}$, S.J. Lilly$^{1}$,
\newauthor P. Capak$^{2}$, Y. Taniguchi$^{3}$, O. Le F\`{e}vre$^{4}$, A. Renzini$^{5}$, N. Scoville$^{2}$,  
\newauthor M. Ajiki$^{6}$, H. Aussel$^{7,8}$, T. Contini$^{9}$,  H. McCracken$^{10,11}$, B. Mobasher$^{12}$, 
\newauthor  T. Murayama$^{6}$,
D. Sanders$^{7}$, S. Sasaki$^{6,13}$, C. Scarlata$^{1}$, M. Scodeggio$^{14}$,
\newauthor Y.  Shioya$^{13}$,  J. Silverman$^{15}$, M. Takahashi$^{6,13}$,  D. Thompson$^{2,16}$, G. Zamorani$^{17}$ 
\\
\\
$^{1}$Institute of Astronomy, Department of Physics, ETH Zurich, CH-8093 Zurich, Switzerland \\
$^{2}$California Institute of Technology, MC 105-24, 1200 East California Boulevard, Pasadena, CA 91125 \\
$^{3}$Subaru Telescope, National Astronomical Observatory of Japan, 650 North Aohoku Place, Hilo, HI 96720 \\
$^{4}$Laboratoire d'Astrophysique de Marseille, BP 8, Traverse du Siphon, 13376 Marseille Cedex 12, France \\
$^{5}$Dipartimento di Astronomia, Universit\'a di Padova,  vicolo dell'Osservatorio 2, I-35122 Padua, Italy \\
$^{6}$Astronomical Institute, Graduate School of Science,
         Tohoku University, Aramaki, Aoba, Sendai 980-8578, Japan \\
$^{7}$Institute for Astronomy, 2680 Woodlawn Dr., University of Hawaii, Honolulu, Hawaii, 96822\\
$^{8}$Service d'Astrophysique, CEA/Saclay, 91191 Gif-sur-Yvette, France\\
$^{9}$Laboratoire d'Astrophysique de l'Observatoire Midi-PyrŽnŽes, Toulouse, France \\
$^{10}$Institut d'Astrophysique de Paris, Universit\`e Pierre et Marie Curie, 75014 Paris, France \\
$^{11}$Observatoire de Paris, LERMA, 61 Avenue de l'Observatoire, 75014 Paris, France\\
$^{12}$Space Telescope Science Institute, 3700 San Martin Drive, MD 21218, USA \\
$^{13}$Physics Department, Graduate School of Science    and Engineering, Ehime University, Japan\\
$^{14}$INAF - IASF  Milano, Italy  \\
$^{15}$Max Planck Institut fŸr Extraterrestrische Physik, Garching, Germany \\
$^{16}$Large Binocular Telescope Observatory, University of Arizona, 
   Tucson, AZ  85721-0065,   USA \\
$^{17}$INAF Osservatorio Astronomico di Bologna,  Italy 
}

\begin{document}  
\date{Submitted: May 8, 2006}
\pagerange{\pageref{firstpage}--\pageref{lastpage}} \pubyear{2006}
\maketitle
\label{firstpage}

\begin{abstract}
\noindent
We present \zebra{}, the Zurich Extragalactic Bayesian Redshift Analyzer. \\ The  current version of  \zebra{} combines and extends several of the classical approaches to produce accurate photometric redshifts down to faint magnitudes. In particular, \zebra{} uses the template-fitting approach to produce  Maximum Likelihood and  Bayesian redshift estimates based on: \\ 
$(1.)$ An automatic iterative technique to correct the original set of galaxy templates to best represent the SEDs of real galaxies at different redshifts;  \\  
$(2.)$ A training set of spectroscopic redshifts for a small fraction of the photometric sample
to improve the robustness of the photometric redshift estimates; and \\ $(3.)$
An iterative technique for  Bayesian redshift estimates, which extracts  the full two-dimensional  redshift  and template probability function for each galaxy. \\  We demonstrate the performance of  \zebra{} by applying it to a sample of  866  $I_{AB}\le 22.5$ COSMOS galaxies with  available $u*$, $B$, $V$, $g'$, $r'$, $i'$, $z'$ and $K_s$ photometry and zCOSMOS spectroscopic redshifts in the range $0 <  z < 1.3$. 
Adopting a 5-$\sigma$-clipping that excludes $\le10$  galaxies, both the Maximum Likelihood and Bayesian  \zebra{}  estimates for this sample have an accuracy $\sigma_{\Delta{}z/(1+z)}$ smaller than 0.03. Similar accuracies are recovered using mock galaxies.\\   \zebra{}  is made available to the public at:  {\it http://www.exp-astro.phys.ethz.ch/ZEBRA.}
\end{abstract}

\begin{keywords}
galaxies: photometric redshifts -- galaxies: distances and redshifts -- galaxies: photometry -- galaxies: formation and evolution -- methods: statistical
\end{keywords}

\section{Introduction}
\label{sec:Introduction}

Current imaging surveys of faint high redshift galaxies such as, e.g., COSMOS \citep{scoville2006}, already return millions of galaxies with magnitudes well beyond the current observational spectroscopic limits. As spectroscopic redshifts for such large distant galaxy samples will thus remain practically unobtainable in the foreseeable future,  photometric redshifts of increasing accuracy will have to be constructed in order to properly exploit the wealth of information, as a function of cosmic epoch,  that is potentially extractable from state-of-the-art and future large imaging surveys.

The importance of estimating accurate redshifts from 
multi-wavelength medium- and broad-band photometry for large galaxy samples is reflected in the extensive efforts that have been devoted to improving algorithms and methodologies to increase the accuracy of the photometric  estimates (see, e.g.,  \citealt{sawicki1997}; \citealt{1998astro.ph..9347Y}; \citealt{arnouts1999}; \citealt{2000ApJ...536..571B} (BPZ); \citealt{2000AJ....120.1588B}; \citealt{firth2003}; \citealt{benitez2004}; \citealt{2006ApJS..162...20B}; \citealt{ilbert2006}, and references therein). These works are based on a few basic principles, namely: 

$(i)$
$\chi^2$ minimization of the difference between a model-galaxy Spectral Energy Distribution (SED; the model SEDs are hereafter referred to as 
{\it templates}) and the observed galaxy photometry; $(ii)$  Neural network approaches that rely on the availability of a small sample of spectroscopic redshifts to 
find a functional dependence between photometric data and redshifts;
$(iii) $
Hybrid approaches that perform  standard $\chi^2$ minimization while using  a small spectroscopic training sample to optimize the initial set of galaxy templates; 
$(iv)$ Bayesian methods which use additional information provided by a prior to obtain final photometric redshift estimates.

Motivated by the scientific returns of deriving accurate photometric redshifts for large numbers of faint COSMOS  galaxies, we have developed \zebra{}, the {\it Zurich Extragalactic Bayesian Redshift Analyzer}.  In this paper we describe the current version of \zebra{}, which 
combines and extends several of the above-mentioned approaches to produce accurate photometric redshifts down to faint magnitudes. More specifically, the paper is structured as follows:

Section \ref{sect:zebraIntro}: {\it About \zebra{}}, specifies the input requirements and the output of \zebra{}.

\begin{figure}
\includegraphics[bb=60 288 571 799,width=84mm]{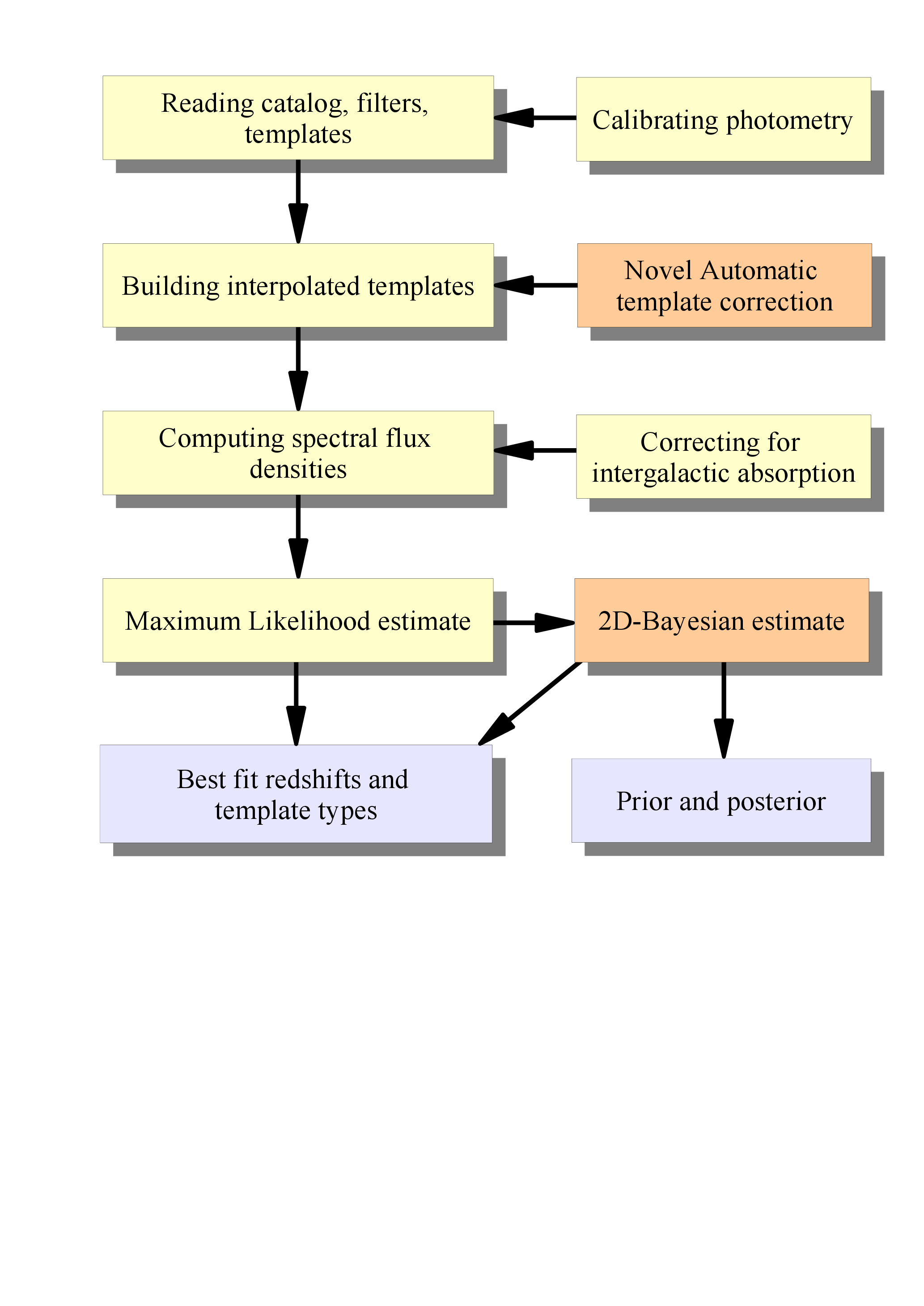}
\caption{The architectural design of \zebra{}: the individual components are described in Section \ref{sect:zebraPrinciples}. The calibration of the photometric catalog and the automatic template corrections are performed  by running \zebra{} in its \pc{} and \teo{} modes, respectively.  
In red are shown the boxes corresponding to the innovative components of \zebra{}; In 
particular the automatic template correction module, and the two-dimensional Bayesian module. The output of \zebra{} is indicated by blue boxes.}
\label{fig:DesignZEBRA}
\end{figure}

Section \ref{sect:zebraPrinciples}: {\it The principles of \zebra{}},
describes  the general design and  methodological details of  the code. A flow chart indicating the architectural structure of  \zebra{} is shown in Figure  \ref{fig:DesignZEBRA}. 
Basically, \zebra{} produces two separate estimates for the photometric redshifts of individual galaxies: A Maximum Likelihood (ML) estimate,  and a Bayesian (BY) estimate. 
These achieve a high accuracy by combining together some novel features with several of the approaches that have been published in the  literature. In particular, \zebra{}:

\begin{description}
\item{-} Uses a novel automatic iterative technique to correct an original set of galaxy templates to best represent the SEDs of real galaxies at different redshifts. These template corrections depend  on the accuracies and systematic errors in the absolute photometric calibrations; therefore,  prior to performing the individual template corrections, \zebra{} automatically removes systematic calibration errors in the input photometric catalogs. The template corrections substantially reduce  the photometric redshift inaccuracies that are generated by galaxy-template mismatches;
\item{-}  Can be fed with a training set of spectroscopic redshifts for a small fraction of the photometric sample, to improve the robustness of the photometric redshift estimates;
\item{-}  Adopts  an  iterative technique for  Bayesian photometric redshift estimates that extracts  the full two-dimensional  redshift  {\it \underline{and}}  template likelihood function for each galaxy.
\end{description}

Section \ref{sect:application}: {\it The $1^{st}$ application of \zebra{}},
demonstrates the performance of  \zebra{} by comparing our photometric redshifts estimates for a sample of $866$   $I_{HST,AB}<22.5$  ACS-selected COSMOS galaxies
with high-quality zCOSMOS  spectroscopic redshifts $z_{spec}\le1.3$ \citep{lil06}.  Based on the currently available passbands and  photometric accuracies,  both the  ML and BY  \zebra{} photo-$z$'s  for COSMOS galaxies have a   $5\sigma-$clipped accuracy of $\Delta{}z /(1+z)=0.027$
over the entire redshift range (with $\sim1\%$ outliers).

Section \ref{sect:Conclusions}: {\it  Concluding remarks},
briefly comments on the first applications of the COSMOS \zebra{} photometric redshifts, and lists  the  developments which we are already implementing in the  next version of  \zebra{}.

The three Appendices
introduce the notation and conventions that we use throughout the paper
(Appendix \ref{sect:definitions}),
present in detail the explicit mathematical formulation of 
\zebra{}'s algorithms (Appendix \ref{sect:tempCorrection}), and demonstrate the \zebra{} performance on a Mock catalogue produced for the COSMOS survey (courtesy of Manfred Kitzbichler; Appendix \ref{sect:mock}).

\zebra{} is made available to the general public at {\it http://www.exp-astro.phys.ethz.ch/ZEBRA}\footnote{The \zebra{} website is currently under construction.}.

The use of \zebra{} should please be acknowledged with an explicit reference  to this paper in the bibliographic list of any resulting publication.

\section{About \lzebra{} }
\label{sect:zebraIntro}

\zebra{} accepts as input:

$(i)$ A  photometric catalog containing medium- and broad-band photometric data for each galaxy of the sample under study;

$(ii)$ The filter transmission curves corresponding to the  passbands of  the photometric catalogue, and

$(iii)$ An initial set of templates.

Optimally, photometric errors should also be included in the photometric catalog; however, it is possible to set errors to a user-specified value.  Some frequently used templates and filter curves are already provided within \zebra{}.

\zebra{} offers a variety of output information depending on the program configuration:

 $-$ When run in \pcm{} (Section \ref{sect:photCheckMode}),   the program corrects the input photometric catalog of any systematic calibration error, and  returns the detailed information about the applied corrections. 

 $-$  In \tom{} (Section \ref{sect:tempCorr}) \zebra{} returns the corrected templates as wavelength versus spectral-flux-density tables. 

 $-$  In  the  \mlm{} (section \ref{sect:ML}),  the main output consists of the best fit redshift and template type for each galaxy in the photometric catalogue, together with their confidence limits estimated from constant $\chi^2$ boundaries. 
Additionally, the program returns: (i) the minimum $\chi^2$, (ii) the normalization of the best fit template, (iii) the rest-frame B-band magnitude and (iv) the luminosity distance. 
If specified by the user, further information is accessible, e.g. the likelihood functions for all galaxies in several output formats, and the residuals between best fit template magnitude and observed  magnitude for each galaxy in each passband.

  $-$  In the \bm{} (Section~\ref{sect:Bayes}), \zebra{} calculates the 2D-prior in redshift and template space in an iterative fashion. This final prior (and, if specified, the interim prior of each iteration step) is provided, together with the posterior for each galaxy. The posterior can be saved as full 2D-table or in marginalized form.  \zebra{}'s output also lists  the most probable redshift and template type for each galaxy, as defined by the maximum of (i) the 2D posterior or (ii) the posterior after marginalizing over templates types and redshifts, respectively. The errors are calculated directly from the posterior. 

  $-$  \zebra{} can also  derive and return  $K$-corrections based on the specified templates and filters.

 All input and output files of \zebra{} are ASCII-files.

Overall, \zebra{} is designed in a flexible way allowing all key-parameters to be user-defined.  A detailed updated description of \zebra{}'s input and output, and a manual explaining its use, can be found at \zebra{}'s URL.

\section{The principles of \lzebra{}}
\label{sect:zebraPrinciples}

\subsection{Step 0: Correction of systematic calibration errors in the input photometric catalogues}
\label{sect:photCheckMode}

In principle, with perfectly calibrated photometry, \zebra{} can be run directly in the \tom{}, so as to determine the optimal corrections to the original templates that allow to properly reproduce the SED of galaxies at all redshifts.  If present, however, systematic calibration errors in the input  photometric catalogs deteriorate the quality of  the photometric redshift estimates.  
Such calibration errors can be easily identified, as they lead to {\it residuals which are independent of template type}  between  best-fit template and  galaxy fluxes.

The \pcm{} of \zebra{} offers the possibility of correcting for  any such possible systematic calibration error before performing any correction to the shape of the individual templates (as done, e.g., in \cite{2000ApJ...536..571B} and \cite{2004AJ....127..180C}; see also \cite{2006astro.ph..5262C} for an application). In particular, the \pcm{}  of \zebra{}:

$-$  Computes, for each galaxy $i$ and for each filter $n$, 
the difference $\Delta{}\text{mag}_{n,i}$ between the magnitude of the best-fit  template and the observed  magnitude $\text{mag}_{n,i}$ of the galaxy in that filter.\footnote{These 
residuals can be calculated using either the full photometric catalogue, or the small
``training set'' of galaxies with spectroscopic redshifts, if available. The latter approach has the advantage that the known redshift can be kept fixed in the template-galaxy fits, thereby reducing the scatter in the detected trends.} 

$-$  Fits, separately for each passband but independent of the template, the dependence of the  $\Delta{}\text{mag}$  residuals on the observed galaxy magnitude. 
A constant shift,  a linear or higher order regression  can be separately applied to each of  the 
$\Delta{}\text{mag}$ vs $\text{mag}$ relations. 

$-$ Applies the derived corrections to each photometric set of data before re-iterating the procedure. 
The photometric corrections clearly depend on the input templates. Hence, it is important to ensure that the initial set of  templates is well adapted to the galaxy types in the catalog and adequately covers the wavelength range which encompasses all passbands at all relevant redshifts. Furthermore, a photometric shift in one passband may lead  to  a change in the normalization of the template fits; Thus, a faster convergence of the iterative procedure can be reached by temporarily increasing the relative error in the specific passband. Tests performed by adding artifical offsets to our photometric data (observed and mock, see Section \ref{sect:dataTemplates} and Appendix \ref{sect:mock}) show that, with this extra step, convergence is always achieved as long as $(i)$ not {\it all} bands need significant photometric corrections (i.e. much larger than the photometric error), and $(ii)$ the passbands are not strongly correlated, e.g. they should not overlap. 
 In Appendix \ref{sect:mock} we further discuss these issues.
 
The main modules of \zebra{} 
are then run on input photometric catalogues that contain no systematic errors in the calibrations.

\subsection{A key element of \zebra: A novel automatic template correction scheme}
\label{sect:tempCorr}

In principle, an advantage of template-matching approaches  for photo-$z$ estimates is that they do not necessarily require  a {\it training set} of galaxies with accurately known redshift  from spectroscopic measurements.  In practice, however, the available templates (e.g., $z=0$ galaxy SEDs or synthetic models) are typically inadequate 
to reproduce the SEDs of real galaxies at all  redshifts.
Therefore, a substantial error in the estimate of the photo-$z$'s in template-matching schemes is contributed by mismatches between real galaxies and available  templates. 

\cite{2000AJ....120.1588B}  propose, as a way to  mitigate this problem, to apply the training set approach within the template-fitting method so as to optimize for the shape of the spectral template that best match  the predicted galaxy colors (calculated using the spectroscopic redshift) and the observed colors. This is done by transforming  the discrete template space into a linear continuous space, and using a  Karhunen-Lo\`{e}ve expansion to iteratively correct, through  a $\chi^2$ minimization scheme, the eigenbases of a lower-dimensional subspace. As a result, spectral templates are derived that are a better match to the SEDs of the galaxies in the training set than are the initial model/empirical templates
(see also \citealt{2000AJ....119...69C}, \citealt{2001AJ....122.1163B}, \citealt{benitez2004}; \citealt{2003AJ....125..580C} present an application of this method to SDSS data).

Given the availability of a training set of galaxies in the redshift interval of interest, \zebra{} uses a 
similar template correction scheme, which however extends and improves on the 
$\chi^2$-minimization approach adopted in the previous works.
The improvements  include:

\begin{enumerate}

\item The simultaneous application of  the minimization scheme to all galaxies in the photometric sample at once;

\item The introduction of a {\it regularization term in the $\chi^2$ expression}, which prevents unphysical, oscillatory wiggles in the wavelength-dependent template correction functions;

\item A formalization of the  $\chi^2$  minimization step that allows the use of  interpolated templates (in magnitude space, so as to better sample the parameter space covered by the available original templates), and includes the effects of intergalactic absorption in a straightforward manner;

\item Template corrections optimized in different user-specified redshift regimes. 

\end{enumerate}

Details on the implementation of the concepts above are given in Appendix 
\ref{sect:tempCorrection}. 
Briefly,  \zebra{} minimizes, 
for all 
catalogue entries $i$ with best fitted template type $t$
at once, the following  $\chi^2$ expression:
\begin{align}
\label{eq:chi2MinimText}
\chi_t^2&=\frac{1}{N_t}\sum_{i=1}^{N_t}\chi_{t,i}^2=\sum_k\frac{1}{\sigma_{t,k}^2}(s_t^{\text{cor}}(k)-s_t^{\text{orig}}(k))^2 \nonumber \\
&+\frac{1}{N_t}\sum_{i=1}^{N_t}\sum_{n=1}^{N_B}\frac{1}{\Delta_{n,i}^2}(f_{n,i}^{\text{cor}}-f_{n,i}^{\text{obs}})^2 \nonumber \\
&+\sum_k\frac{1}{\rho_{t,k}^2}(s_t^{\text{cor}}(k+1)-s_t^{\text{cor}}(k)-s_t^{\text{orig}}(k+1)+s_t^{\text{orig}}(k))^2,
\end{align}
with the following definitions:

$-$  $N_t$ is the set of catalogue entries, and contains all galaxies which are best fitted by template type $t$.

$-$ $\sigma_{t,k}$  is a  {\it pliantness parameter}  that regulates the amplitude of the deviations of the corrected template shape from the initial template shape.

$-$ $s_t^{\text{cor}}(k)$ is the corrected template shape for template type $t$, and is obviously  a function of the wavelength $k$.

$-$ $s_t^{\text{orig}}(k)$ is the shape of the original template $t$.

 $-$ ${\Delta_{n,i}}$ is the error of the photometric flux density in  filter $n$ for galaxy $i$.

$-$  $f_{n,i}^{\text{cor}}$ is  the spectral flux density of the {\it corrected} best fit template $t$ in filter band $n$ for galaxy $i$. The dependence on the  best fit template type $t$, best fit redshift $z$ and template normalization $a$ is left implicit.

 $-$  $f_{n,i}^{\text{obs}}$ is the observed spectral flux density of  galaxy $i$ in filter band $n$.

 $-$  ${\rho_{t,k}}$ is the {\it regularization parameter}, which constrains the gradients between original and corrected template shapes. The smaller $\rho$, the stronger the suppression of high-frequency oscillations in the shape of the corrected templates.

The second  term 
in the r.h.s. of equation \eqref{eq:chi2MinimText}
minimizes the difference between observed flux $f_{n,i}^{\text{obs}}$ and template flux $f_{n,i}$ for all passbands $n$, averaged over all galaxies $i$. With the appropriate choice for the pliantness and regularization parameters, the first and third terms ensure that the correction procedure generates only templates with physically acceptable shapes. Specifically, the first  term prevents too large deviations between the corrected and  the uncorrected  templates, and the last term regularizes the shape and inhibits strong oscillations in the SED of the corrected templates.
Therefore, the minimization of the so-defined $\chi_t^2$  is a compromise between two orthogonal requirements:  On the one hand, each original template is changed so that, averaging over all galaxies $i$ which are best fitted by that given original template, the corrected spectral flux density
closely matches the measured spectral flux density. On the other hand, unphysical, large oscillations over small wavelength ranges 
are avoided when correcting the shape of the templates.
The self-regulation terms maximize the stability and reliability of the template corrections , especially when only a small training and/or  modest S/N   data are available. 

\begin{figure}
\begin{minipage}{80mm}
\includegraphics[width=80mm]{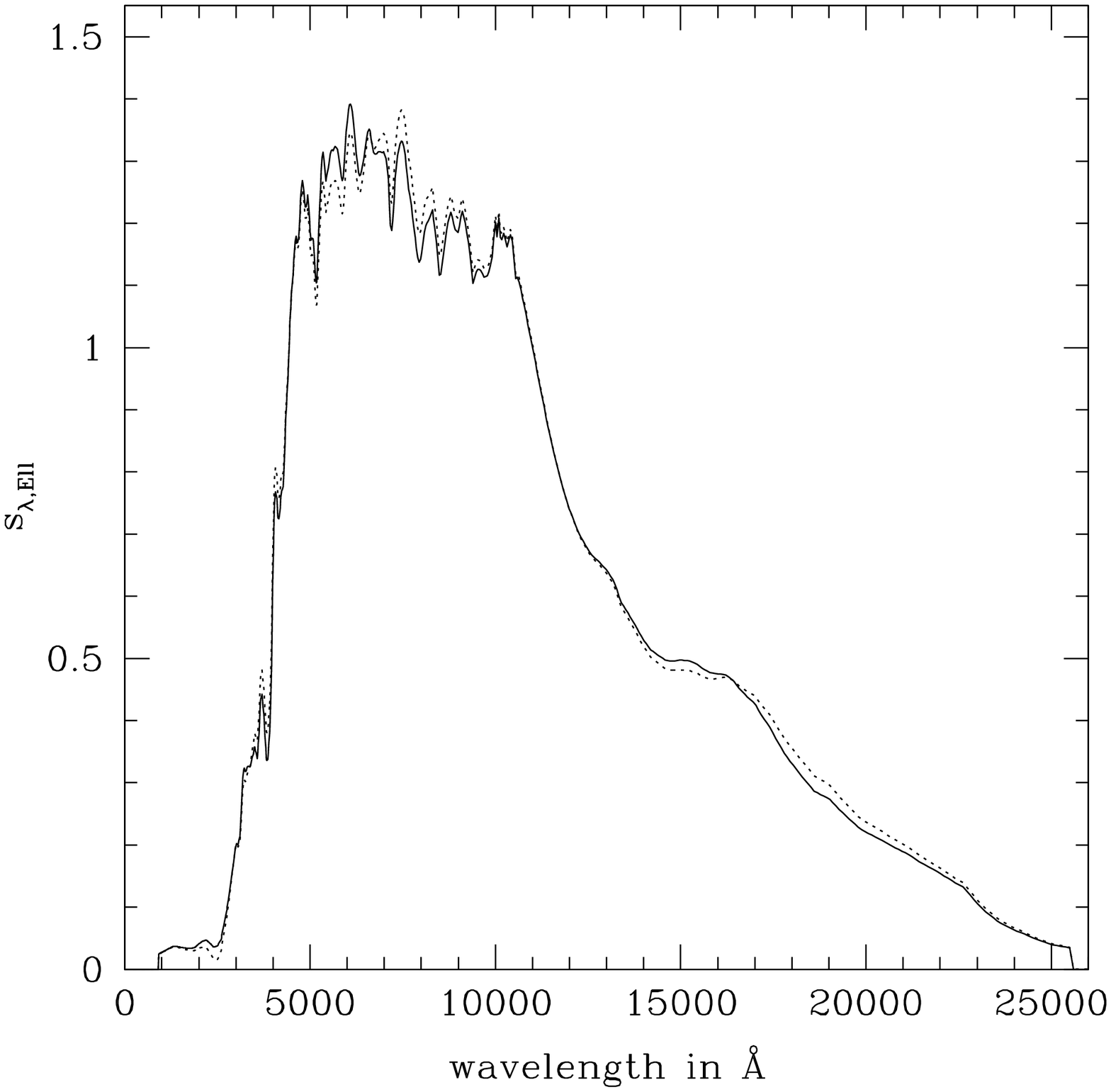}
\end{minipage}
\begin{minipage}{80mm}
\includegraphics[width=80mm]{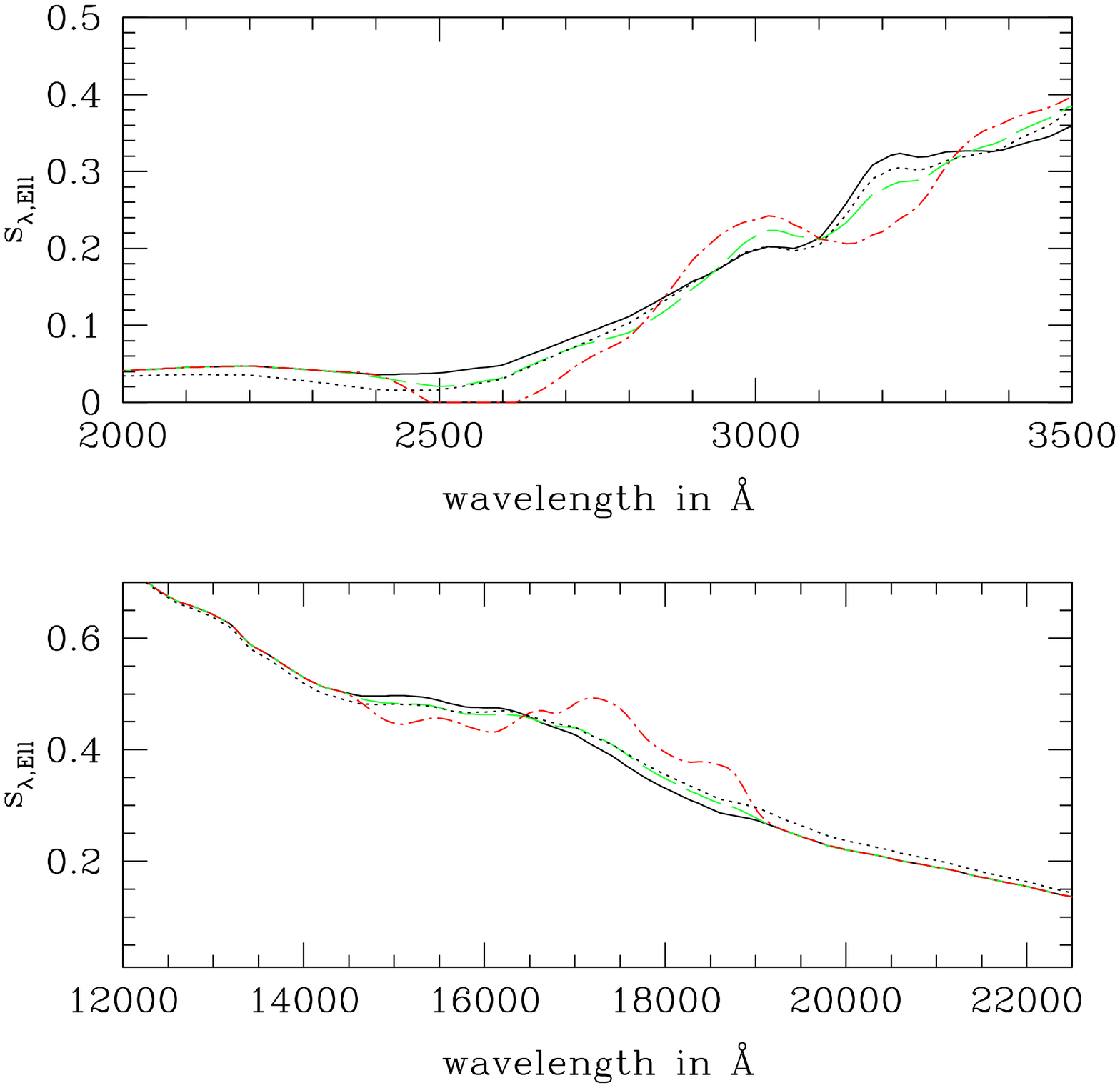}
\end{minipage}
\caption{
The figure shows the  effect of varying the pliantness $\sigma$ and the regularization parameter $\rho$ on the shape of the elliptical template. Top: The original template of an elliptical galaxy (solid black line) is compared with the corrected template using the values $\sigma=2$ and $\rho=0.05$ (dotted black line). 
Middle and Bottom: The results of using different parameter choices are shown in detail. The solid and dotted line correspond to the same templates as above. The dashed-green ($\sigma=0.4$) and dot-dashed red ($\sigma=2$) lines result from applying the template correction scheme without regularization, i.e. setting $\rho=\infty$.
The unregularized template with $\sigma=0.4$ is well-behaved, but this low value of $\sigma$ is still inadequate to properly fit the SEDs of observed galaxies. However, when choosing a five times higher pliantness ($\sigma=2$) to try to improve the correction, strong unphysical oscillations develop in wavelength ranges that are smaller than the width of the filters. The template changes are  localized  in separated wavelength regions and lead to unrealistic, distinguished bumps in the template shape. The high-frequency oscillations may even require to set the flux of the corrected template to zero, in order to avoid negative spectral flux densities. These unphysical ``over-corrections''  are avoided by choosing a finite regularization parameter $\rho$.}
\label{fig:varyingSigma}
\end{figure}

 In principle, the optimal values of  $\sigma$ and $\rho$ might be both template and wavelength dependent. In Figure~\ref{fig:varyingSigma} we show the effects of varying $\sigma$ and $\rho$ in the template correction procedure.  In particular, a too small value for $\sigma$ inhibits  template changes and thus reduces the efficacy 
of the corrections, and a too large value for $\sigma$ leads to  unphysical high-frequency oscillations in the shape of the corrected templates. The latter effect can 
be avoided by choosing an appropriate value for $\rho$.

The \zebra{} template correction is implemented in  two main steps.
The procedure is started by using in step 1 only  the original templates, but is  iterated so that each new iteration of step 1 uses the combined set of original and corrected templates. The two main steps are: 

$(1.)$  Computation of the set ${N}_t$ that contains  all galaxies which are best fitted by the template $t$ or (from second iteration on) by a corrected template originating from template $t$.

$(2.)$ Correction  of the template shape of each {\it original}  template $t$ by minimizing the corresponding $\chi_t$ expression.

The  two steps are repeated several times, as the  best fit template type  might change when considering new  corrected templates in step 1 in the computation of ${N}_t$.

We note that \zebra{} can perform logarithmic interpolations of the original (and corrected) templates; thus, the  $\chi^2$ that is actually minimized in the code is modified relative to the expression above so as to take this into account (see Appendix B for details).  

Finally, \zebra{} can optimize the automatic template  corrections in different redshift ranges by 
grouping  the catalog entries in different redshift bins before the $\chi^2$ minimization step. This option, tested on the COSMOS data (Section 4), is found to substantially improve the reliability and quality of the \zebra{} template corrections.  
In Appendix \ref{sect:mock} we test the method further by applying it to a mock catalog for the COSMOS survey.

\subsection{The \zebra{} Maximum Likelihood  module}
\label{sect:ML}

The estimation of ML redshifts constitutes the core of the \zebra{} code.  
To produce the ML redshifts, \zebra{} performs the following steps (see Figure \ref{fig:DesignZEBRA}):

\begin{enumerate}

\item Read the input photometric catalog, filters and templates. The data in the
catalogs (expected to be in magnitudes) are converted into spectral flux densities. 
Data errors are either  read from the catalog or specified by the user 
in one of several formats.

\item Interpolate  the original templates (optional). Two interpolation schemes are implemented, namely  
interpolation in magnitude space (``log-interpolation'') and in spectral flux density (``lin-interpolation''); these can also be used simultaneously. Specifically,  a set of original templates 
is first sampled on a fixed wavelength grid, and then used to define the log-interpolated 
templates
$s^{\text{log}}_{t_1,t_2,g}(k)$
with the weight $g$ relative to the  two basic adjacent templates $s_{t_1}(k)$ and  $s_{t_2}(k)$ defined as:
\beq
\label{eq:defLogInterpol}
s^{\text{log}}_{t_1,t_2,g}(k)=(s_{t_1}(k))^{1-g}(s_{t_2}(k))^{g},\quad{}g\in(0,1).
\eeq
The lin-interpolated templates  $s^{\text{lin}}_{t_1,t_2,g}(k)$ are instead linear combinations of the basic
templates:
\beq
s^{\text{lin}}_{t_1,t_2,g}(k)=(1-g)\thinspace{}s_{t_1}(k)+g\thinspace{}s_{t_2}(k),\quad{}g\in(0,1).
\eeq

\item Sample the filter curves  on two different grids. The first grid is equal to the one used for the templates, and coarsely samples the filter shapes (as most of its elements are  in wavelength bins where the transmission of the filters are equal to zero); this grid is used to optimize the speed of \zebra{} within the template correction scheme. The second grid is optimized to sample with high accuracy each individual filter in its transmission window; this high-resolution grid is used to calculate the spectral  flux densities for each template in the different filter bands. 

\item Correct the filter transmission functions  for sharp features occurring at particular wavelengths by smoothing with  a top-hat kernel. These modifications to the original filter curves are found to prevent artificial peaks in the likelihood functions; these peaks arise when e.g., a strong emission line in the galaxy or template spectrum is ``trapped'' in a filter-transmission ``hole'', returning an overall spuriously small $\chi^2$ value. 

\item Calculate the (redshift-optimized) corrections for  each  original templates  as described in Section~\ref{sect:tempCorr}, and extend the set
of available templates to include both the original and the corrected templates, and their interpolations.

\item Calculate, for each template $t$ and redshift $z$, the spectral flux densities $f_{z,t,n}$  in each filter band $n$. The mean optical depth $\tau(\lambda,z)$ of the intergalactic absorption is computed according to either \cite{1995ApJ...441...18M} or \cite{2006MNRAS.365..807M} (see Appendix \ref{sect:definitions}). The $f_{z,t,n}$ values are stored in a three-dimensional array.

\item Determine the best fitting template normalization factor $a^*(z,t)$ and the value of $\chi^2(z,t,a^*)$  using  the formulation of \cite{2000ApJ...536..571B} (see Equation \eqref{eq:Chi2Benitez}, Appendix \ref{sect:definitions}). A search in the two-dimensional array
$\chi^2(z,t,a^*)$ is carried out to find the minimum $\chi^2$ and thus the best fitting values $z^*$
and $t^*$.  A pair $(z,t)$ is accepted only if the absolute $B$ magnitude $M_B$  lies within some (user-supplied) limits, so as  to avoid mathematically good fits which are however physically unacceptable (as they would imply unrealistically  dim or bright galaxies at a given redshift). Similar constraints are adopted by other authors, e.g.,  \cite{2003MNRAS.345..819R} adopt  the range  $-22.5<M_B<-13$. 

\item Calculate the errors on the ML best fit redshift estimates   using constant boundaries $\chi^2_{\text{min}}+\Delta{}\chi^2$ as (two-parameter) confidence  limits. For  Gaussian-distributed errors $\Delta_n$, the values $\Delta{}\chi^2$=2.3 and $\Delta{}\chi^2$=6.17  correspond to 1-$\sigma$ and 2-$\sigma$ confidence limits, respectively. This means that the probability that the ``true'' value pair $(z^{\text{true}},t^{\text{true}})$ falls in an elliptical region which extends
within $[z^*-\Delta{}z,z^*+\Delta{}z]$ when projected to the $z$ axis, and within 
$[t^*-\Delta{}t,t^*+\Delta{}t]$ when projected to the $t$ axis,  is  $68.3\%$ and $95.4\%$, respectively.

\end{enumerate}

The  \zebra{}'s ML module computes the full likelihood functions in the two-dimensional redshift-template space, which are then used as input for the \zebra{} BY estimates.

\subsection{The \zebra{} two-dimensional Bayesian  module}
\label{sect:Bayes}

As discussed  in \cite{2000ApJ...536..571B} and \cite{2006ApJS..162...20B}, employing the Bayesian method for the determination of photometric
redshifts enables the inclusion of  prior knowledge on the statistical properties of the galaxy sample under study, and thus to substantially improve, statistically, the accuracy of the redshift estimates. 

The general idea behind  Bayes theorem is that the ``posterior'' $P(\boldsymbol{\alpha}|\mathbf{f})$, which 
provides the parameters 
$\boldsymbol{\alpha}$ given the data $\mathbf{f}$,  can be determined if the ``prior'' $P(\boldsymbol{\alpha})$ and  the likelihood $\LL(\boldsymbol{\alpha})$ are known. Specifically:
\beq
P(\boldsymbol{\alpha}|\mathbf{f})=P(\boldsymbol{\alpha})\frac{\LL(\boldsymbol{\alpha})}{P(\mathbf{f})}.
\label{eq:PriorAbstract}
\eeq

Despite its name,  the function $P(\boldsymbol{\alpha})$ might 
not be known a priori. 

In \cite{2000ApJ...536..571B}, the Bayesian prior is calculated by assuming an analytic function and fixing its free parameters using the available galaxy catalog. The method is  powerful: an application is presented in \cite{benitez2004}. In particular, by construction, the resulting  redshift distribution is  smooth and the effects of cosmic variance are reduced. The chosen analytic form may however not necessarily take properly into account the selection criteria of the galaxy catalog under study; furthermore,  some assumptions on the relation amongst the different free parameters are required in order to constrain the fit.

A different approach is described in \cite{2005MNRAS.359..237P}. There, the true redshift density distribution is estimated by "deconvolving" the measured maximum-likelihood redshift distribution from the errors of the photometric redshift estimates.  This method has the advantage of being very general;  however, for degenerate distributions and/or a small galaxy samples, it may not converge to a stable solution unless an additional prior is introduced. 

To address these issues, \cite{2006ApJS..162...20B} propose an iterative method to build the prior self-consistently, using as a start the input photometric catalogue; in the redshift domain, this method has the advantage of closely matching  specific  over- and under-densities in the redshift distribution of the target field which are due to cosmic variance. These authors present extensive tests, performed on the galaxy data and using Monte Carlo simulations, to show that the method converges to a stable prior.

\zebra{} adopts the same self-consistent  technique used by \cite{2006ApJS..162...20B} to derive Bayesian estimates for galaxy photometric redshifts, and furthermore extends that formulation by  applying the Bayesian analysis to the full two-dimensional space of 
redshift {\it \underline{and}} template. The equation for the prior is therefore re-written as:
\beq
P(z,t|\mathbf{f}_i^\text{obs})=P(z,t)\frac{\LL_i(z,t)}{\sum_{z,t}P(z,t)\LL_i(z,t)}.
\label{eq:Prior1}
\eeq

Naturally, the so-constructed prior will depend on the selection
criteria for the input sample.  This dependence is carried over into the posterior 
$P(z,t|\mathbf{f}^\text{obs})$, which therefore represents the probability density of determining the 
correct  $z$ and $t$, given the 
observed flux densities
$\mathbf{f}_i^\text{obs}$ {\it and}  the selection criterion for the sample. 
Also, note that the values $z^*$ and $t^*$ of the maximum likelihood solution, and the values $z^\#$ and $t^\#$ which maximize  the posterior probability, are generally different, as the latter are weighed by  the prior. 

The prior is determined by starting with an user-specified 
 guess-prior 
$P_\text{old}(z,t)$ (e.g. a flat prior; as long as the 
initial guess 
is smooth enough, the iterative prior calculation converges quickly to a unique answer; see Section \ref{sect:smoothPrior}), and calculating an improved 
prior $P_\text{new}$
as:
\begin{align}
\label{eq:PriorIter}
P_\text{new}(z,t)&=P_\text{old}(z,t)\frac{1}{N_G}\sum_{i=1}^{N_G}\frac{\LL_i(z,t)}{\sum_{z',t'}P_\text{old}(z',t')\LL_i(z',t')} 
\end{align}
Equation \eqref{eq:PriorIter} follows from 
equation \eqref{eq:Prior1} by assuming that the sample is large enough to be representative, i.e.
\begin{align}
\frac{1}{N_G}\sum_{i=1}^{N_G}P(z,t|\mathbf{f}_i^\text{obs})\approx{}P(z,t),
\end{align}
with $N_G$ the number of galaxies in the sample. By constraining the prior to remain smooth at each iterative step (by convolution with a Gaussian kernel; see below), a small number of iterations, performed by resetting, after each iteration, $P_\text{old}(z,t)\leftarrow{}P_\text{new}(z,t)$, are found to  converge to a stable result for the final prior 
$P(z,t)$.

In practice, it is clearly advisable to exclude unreliable redshift determinations
 in the calculation of  equation \eqref{eq:PriorIter}; these can be contributed by  galaxies with poor template fits and 
 by galaxies with too sparsely sampled SEDs (i.e., with photometric data  in only a small number of passbands).  
In our application of  \zebra{} to the COSMOS data (Section \ref{sect:application}), we define as ``good fits'' those with values of
 $\chi^2$  smaller than the threshold  $\chi^2_\text{0.99}$; this threshold is defined by the condition that, assuming
 that the $\chi^2$ values follow a $\chi^2$ distribution with $N_{filter}-3$ degrees of freedom, the 
 probability of having a value of $\chi^2$ smaller than $\chi^2_\text{0.99}$ is 99\%.
For example, for our special application to the COSMOS data with photometry in eight filters (i.e., for five degrees of freedom),  the threshold  is given by $\chi^2_\text{0.99}\approx{}15$. We have tested, using as thresholds also some specified percentiles of the measured $\chi^2$ distribution, that the final result is rather insensitive to the choice of the threshold. 

\subsubsection{Smoothing of the prior}
\label{sect:smoothPrior}

In principle, the probability density distribution of finding a galaxy
at a given redshift should be  a smooth function of $z$. In practice, however, 
$N(z)$ is estimated from the  galaxy survey under study. The biased sampling of the large-scale structure,
due to the finite area covered by the specific survey, and the shot-noise,
due to the finite number of galaxies in the survey, generate high-frequency 
fluctuations in the observed redshift distribution.
The presence of sharp features in the estimated number counts leads to
a runaway effect in the iterative procedure to determine the best prior.
For galaxies whose likelihood peaks close to the redshift of these features,
the redshift estimation is fully driven by the prior.
Therefore, peaks in the number counts become more and more prominent after
every iteration at the expenses of the surrounding regions.
The net effect is that, after a few iterations, the prior becomes very
spiky.

This instability needs to be eliminated  for a proper Bayesian estimation of galaxy
redshifts. This can be done by building on the key ideas for introducing a prior, which
are $(a)$ to account for the fact that all redshifts are not
equally likely,  and $(b)$ to help to distinguish between degenerate peaks of
the likelihood functions. Therefore, the prior should not 
contain features that  are narrower than the characteristic width of the peaks in
the likelihood functions.

\begin{figure}
\begin{minipage}{80mm}
\includegraphics[width=80mm]{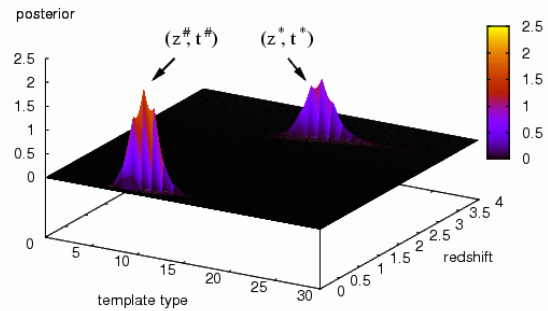}
\end{minipage}
\caption{The two-dimensional probability distribution in template and redshift space for one of the COSMOS galaxies in the sample that we discuss in Section~\ref{sect:application}. For illustration purposes,  the corrected templates used for this run of \zebra{} are collapsed in the figure onto the corresponding 31 ``uncorrected'' (original plus log-interpolated) templates. The values of $z^*$ and $t^*$ of the Maximum Likelihood solution, and the values $z^\#$ and $t^\#$ which maximize  the posterior probability, are labeled in the figure.}
\label{fig:examp}
\end{figure}

A simple way to solve the problem is to smooth the prior after each iteration\footnote{Equivalently,
one can smooth the likelihood functions as in \cite{fernandez2002} and \cite{2006ApJS..162...20B}.}.
The smoothing scale must be chosen by comparing a number of
characteristic scales:

$(a)$ The intrinsic broadness (in redshift space) of the features originated
by large-scale structures, $\sigma_{\rm LSS}$;

$(b)$  The standard error of the Maximum Likelihood estimator,
$\sigma_{\rm ML}$;

$(c)$  The typical broadness of the likelihood functions, $\sigma_{\cal L}$
(which, when photometric errors are properly estimated, has to be comparable
with $\sigma_{\rm ML}$); and

$(d)$ The characteristic scale of the oscillations due to finite Poisson
sampling, $\sigma_{\rm P}$ (basically
the maximum redshift difference between two Maximum Likelihood estimates
with consecutive redshifts).

\begin{figure}
\begin{minipage}{80mm}
\includegraphics[width=80mm]{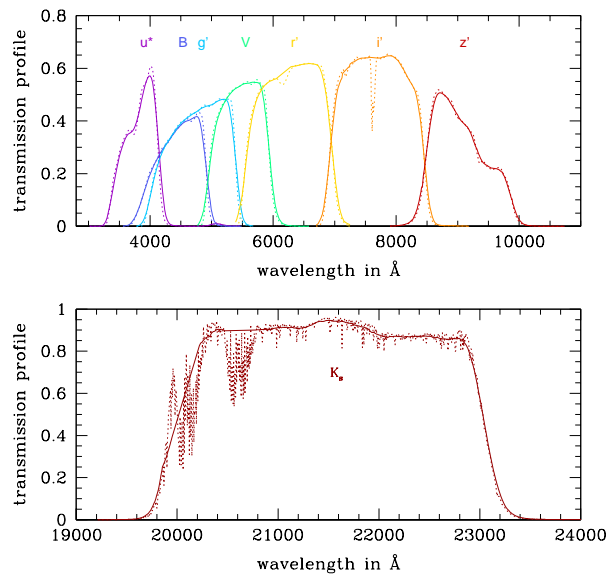}
\end{minipage}
\caption{The transmission curves for the eight filters used to derive the COSMOS photo-$z$'s discussed in this paper. The original filters are shown as dotted lines. The adopted filter shapes are shown as solid lines. We removed ``holes'' in the transmission curves and smoothed them using a top-hat kernel with a FWHM of 200 \AA.
Top panel: the COSMOS filters $u^*$, $B$, $g'$, $V$, $r'$, $i'$ and $z'$. Bottom panel: The original and adopted $K_s$ filter shapes.}
\label{fig:filts}
\end{figure}

We have studied the effect of these different sources of error
by performing a series of Monte Carlo simulations. In brief, we first 
Poisson-sample a given redshift distribution with -- and without -- sharp features
generated by large-scale structures,  and then apply our iterative
procedure assuming Gaussian-shaped likelihoods.
Convergence to a smooth prior is always achieved, in a few iterations,
by smoothing the number counts with a Gaussian kernel of width
$\sigma={\rm max}(\sigma_{\rm ML},\sigma_{\cal L},\sigma_{\rm P})$.
Note that, at low-redshifts, where both $\sigma_{\rm ML}$ and
$\sigma_{\cal L}$ are small,  and for large samples, where also
$\sigma_{\rm P}$ is small, the prior might be affected by
the presence of large-scale structures.
Basically all features such that $\sigma_{\rm LSS}>\sigma$
are broad enough to be robustly detected and
are present in the final prior distribution.
This enhances the probability of measuring redshifts close to e.g.,
the location of large overdensities, and
leads to an optimal estimation of photometric redshift in a galaxy survey.

In \zebra{} we have thus implemented a routine to smooth the prior,  at each step of the iterative procedure described above, by convolution with a Gaussian 
kernel with a user-specified sigma.

\subsubsection{The two-dimensional probability distribution in redshift and template space}

As an example, in Figure~\ref{fig:examp} we show the two-dimensional probability distribution in template and redshift space for one of the COSMOS galaxies in the sample that we discuss in Section~\ref{sect:application}. The values of $z^*$ and $t^*$ of the Maximum Likelihood solution, and the values $z^\#$ and $t^\#$ which maximize  the posterior probability, are indicated in the figure. The distribution shows multiple peaks, and is dramatically different from e.g. the  Gaussian shape that would be typically associated with  a ML photometric redshift estimate. The key strength of the Bayesian analysis is indeed to provide, for each galaxy in a sample,  such detailed information, as this is crucial to almost all statistical analyses of the evolution of galaxy properties with redshift.  

\begin{figure}
\begin{minipage}{80mm}
\includegraphics[bb=15 140 564 690, width=80mm]{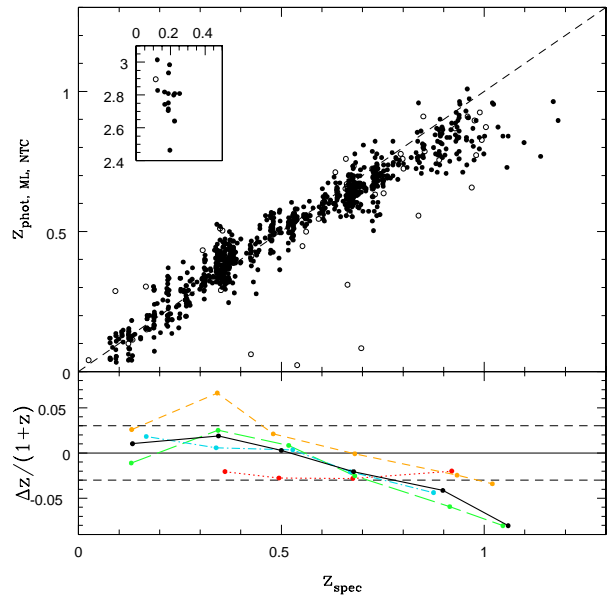}
\end{minipage} 
\caption{
 ML photometric redshifts for the COSMOS sample under study, derived from the \emph{uncorrected} \citet{1980ApJS...43..393C} and \citet{1996ApJ...467...38K} templates plus their log-interpolated templates. In the upper panel, the redshift estimates  $z_{\text{phot,ML,NTC}}$ are plotted against the zCOSMOS $z_{\text{spec}}$ spectroscopic redshifts. Each symbol in the plot corresponds to an individual galaxy. Empty symbols indicate a ``bad'' fit, defined as a fit with  a reduced $\chi^2>$3. The lower panel shows the dependence of the accuracy of the photometric estimates, as quantified by $\Delta{}z/(1+z)$ (with $\Delta{}z=z_{\text{phot,ML,NTC}}-z_{\text{spec}}$), as a function of  $z_{\text{spec}}$. Colors represent different templates: elliptical (dotted red), Sbc (short-dashed orange), Scd (long-dashed green) and irregular (dot-dashed blue) types. The total residual, independent of template type, is shown by a solid black line.
Only templates which contain at least five objects in the respective redshift bin are shown. Interpolated template types are rounded to their nearest basic template type and plotted with the corresponding color. The short-dashed $\Delta{}z/(1+z)\pm{}0.03$ lines correspond roughly to 1-$\sigma$ error bars and are shown to guide the eye.}
\label{fig:ResTempNoImp}
\end{figure}

\section{The $1^{st}$ application of \lzebra{}: z-COSMOS-trained redshifts for COSMOS}
\label{sect:application}

\subsection{The data, the sample and the input  templates}
\label{sect:dataTemplates}

A detailed comparison of zebra{}'s photo-$z$ estimates with those obtained with other codes is presented in Mobasher et al.\ (2006). Here we limit the  demonstration of the performance of \zebra{} by using a sample of 866  $z<1.3$,  $I_{AB} \le 22.5$ COSMOS galaxies with currently available accurate (i.e., ``confidence class''  3 and 4) spectroscopic redshifts  from  zCOSMOS  (the ESO VLT spectroscopic redshift survey of the  COSMOS field; \citealt{lil06}).  A further test on mock galaxies is presented in Appendix \ref{sect:mock}.

These 866 galaxies with zCOSMOS spectroscopic redshifts belong to the complete sample of about $55000$ $I_{AB}\le{}24$ COSMOS galaxies  discussed in \cite{scarlata2006}; we use this complete sample to construct the initial guess-prior in the Bayesian calculation of the photometric redshifts for the COSMOS galaxies.
The allowed range for the galaxy absolute $B$ magnitudes was conservatively fixed to be $-24<M_B<-13$.

\begin{figure}
\begin{minipage}{80mm}
\includegraphics[width=80mm]{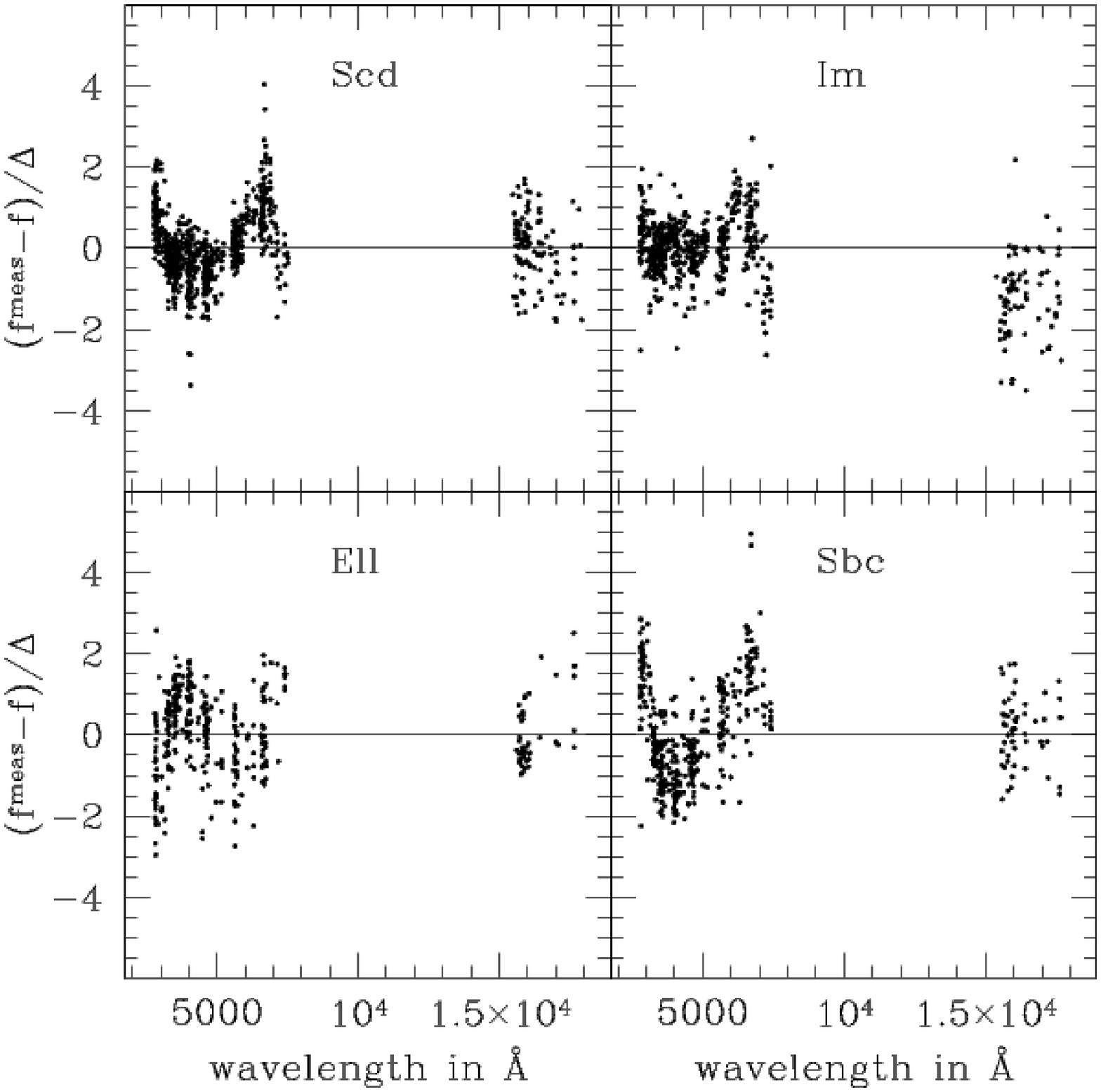}
\end{minipage}
\begin{minipage}{80mm}
\includegraphics[width=80mm]{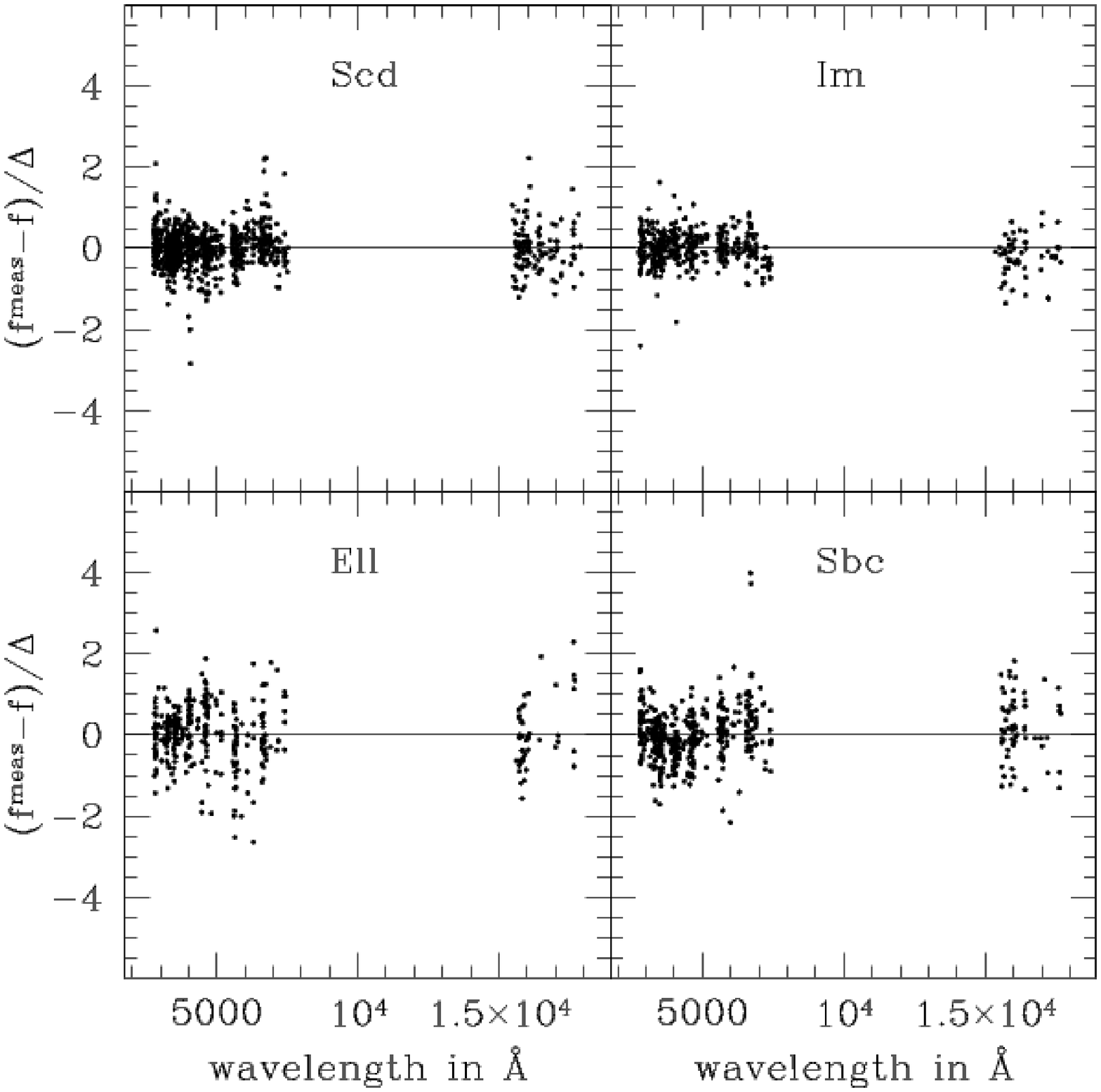}
\end{minipage}
\caption{Error-normalized flux residuals $(f_{n,i}^{\text{obs}}-f_{n,i})/\Delta_{n,i}$ versus rest-frame wavelength for 269 galaxies with spectroscopic redshifts in the range $0.2\leq{}z<0.4$. The different panels correspond to different templates: elliptical (``Ell'') types, Sbc types, Scd types and irregular (``Im'') template types.
 Top: Residuals before the \zebra{}'s automatic template correction is applied. Bottom: Residuals after \zebra{}'s template correction. The systematic trends and scatter are substantially reduced.}
\label{fig:FluxResiduals}
\end{figure}

Exploiting the wealth of ancillary data that are available for the entire COSMOS field, we use, as input photometric catalogues, Subaru  $B$, $V$, $g'$, $r'$, $i'$ and $z'$ photometry ($5\sigma$ magnitude limit of $\sim$27 for point sources in all bands; \citealt{taniguchi2006});  CFHT
$u^*$ photometry ($5\sigma$ magnitude limits for point sources of
$u^*=27.4$); and  $K_s$ photometry from the NOAO wide--field IR imager Flamingos
(Kitt Peak 4m telescope) and the Cerro Tololo ISPI  (Blanco 4m telescope; all data are collected in the catalogue presented by \citealt{capak2006}).  About 
97.2\% of our spectroscopic sample has photometry available in all eight passbands.Ê The relevant  filter transmission curves are shown in Figure~\ref{fig:filts}, before and after the correction for sharp features at specific wavelengths. Systematic calibration errors in each passband were estimated and corrected using the \pc{} module of \zebra. These were typically very small, constant shifts. The robustness of the corrections  was tested by deriving them also after  fixing the redshift  to the spectroscopic value in the galaxy-template fits. 

The adopted  ``original'' set of templates 
consists of the six templates described in  \cite{2000ApJ...536..571B} (which are available on the BPZ website).  These are based on six observed galaxy spectra, i.e., the four of \citet{1980ApJS...43..393C}, i.e., an elliptical, a Sbc, a Scd and an irregular type template,  and the two starbursting galaxy  spectra of \citet{1996ApJ...467...38K}.
\cite{sawicki1997},  \cite{1999ApJ...515L..65B} and \cite{2000AAS...19711704Y} discuss and demonstrate the improvements in the quality of the redshift estimates that are obtained by augmenting the set of templates to include the starbursting types. As discussed by these previous authors, these observed templates are extended into the UV by means of  a linear extrapolation up to the Lyman-Break, and into the IR (up to $\sim$ 25000\AA) using GISSEL synthetic templates.
We furthermore performed a  5-step log-interpolation to sample more densely the SED space covered by the original templates. This results in  a basic set of  31 input  (``uncorrected'') templates.

\subsection{Results}

To illustrate the importance of the \zebra{} template correction, we first present the comparison between the ML photometric redshifts derived when  {\it no template correction} is performed ($z_{\text{phot,ML,NTC}}$), and the zCOSMOS spectroscopic redshifts ($z_{\text{spec}}$).   Figure \ref{fig:ResTempNoImp} presents this comparison.

In the figure, the bottom panel  shows the deviation $\Delta{}z/(1+z)$ versus $z_{\text{spec}}$ (with $\Delta{}z=z_{\text{phot,ML,NTC}}-z_{\text{spec}}$), color-coded for the different templates types.  Although over the entire $0\leq{}z_{\text{spec}}<1.3$ range the overall redshift estimate is acceptable
(a $5-\sigma$-clipped $\sigma_{\Delta{}z/(1+z)}=0.043$ with 19 clipped galaxies),  the  individual templates show large  systematic deviations. The elliptical and Sbc templates in particular show a significant systematic under- and overestimation of the redshifts, respectively. Furthermore, no available ``uncorrected''  (i.e., original plus log-interpolated) template appears to be  adequate to reproduce the SEDs of $z>0.8$ galaxies: these high redshifts are systematically underestimated when using the available z=0 galaxy templates.  While it remains to establish whether this systematic failure at $z>0.8$ is due to the uncertainties  in the templates or to astrophysical reasons (e.g.,  much stronger emission lines at high redshifts than at $z=0$, or  a young, passively evolving elliptical galaxy population; etc), it is clear that  this  systematic effect  would have  a substantial impact on the reliability of statistical studies  of  galaxy evolution with redshift.

\begin{figure}
\begin{minipage}{80mm}
\includegraphics[bb=15 140 564 690, width=80mm]{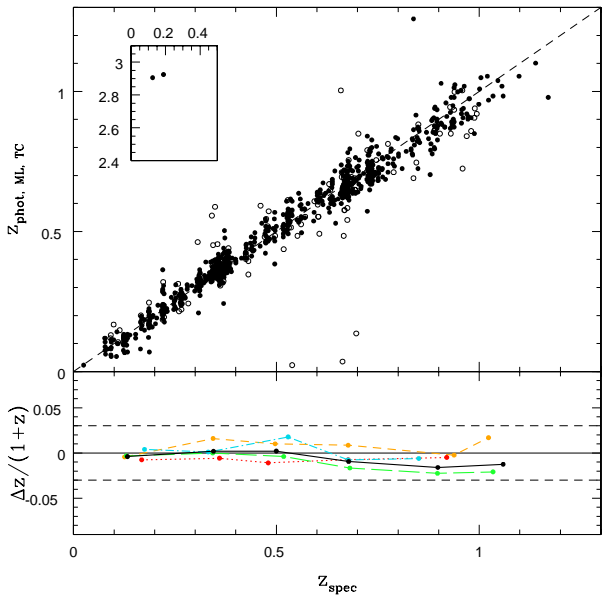}
\end{minipage} 
\caption{
\zebra{}'s ML redshift estimates for the spectroscopic sample derived \emph{after} correcting the  templates as described in Sections \ref{sect:tempCorr} and \ref{sect:application}. The displayed quantities are the same as in Fig.~\ref{fig:ResTempNoImp}. The systematic trend that is visible in Fig.~\ref{fig:ResTempNoImp}, i.e. the underestimation of the redshifts for $z \ge 0.8$, is here eliminated by the use of adequately corrected templates.}
\label{fig:ResTempImp}
\end{figure}

The template correction substantially  improves the photometric redshift estimates, and in particular cures the most troublesome systematic failures of the estimates derived without template correction. As an example, for galaxies with redshifts in the range $0.2\leq{}z<0.4$, Figure \ref{fig:FluxResiduals}    shows the residuals $\Delta{}f$
between observed flux density and best-fit template flux density,
as a function of rest-frame wavelength, before and after template corrections (using $\sigma=2$ and  $\rho=0.05$\footnote{A large volume of $\sigma$-$\rho$ parameter space was explored. Tests show that the \zebra{} solutions are quite stable and do not depend on small variations of these parameters.}).  The substantial improvement in the redshift estimates  is  observable in Figure~\ref{fig:ResTempImp}, which  shows the same comparison with the zCOSMOS spectroscopic redshifts  as above, but
this time for the \zebra{} photometric redshift estimates {\it with template correction}  ($z=z_{\text{phot,ML,TC}}$). The template corrections  were optimized in the redshift bins $z$=0-0.2; 0.2-0.4; 0.4-0.6; 0.6-0.8; 0.8-1.0; 1.0-1.3  and 0-0.3; 0.3-0.5; 0.5-0.7; 0.7-0.9; 0.9-1.3\footnote{The choice of overlapping redshift bins was made to avoid spurious  ``boundary'' effects in the derivation of the redshift-optimized templates.}). Note that the $z_{\text{phot,ML,TC}}$ redshifts at and above $\sim0.8$ lie now well within the statistical errors. The global accuracy of the \zebra{} ML redshift estimates $z_{\text{phot,ML,TC}}$ is  now reduced to a 5-$\sigma$-clipped $\sigma_{\Delta{}z/(1+z)}=0.027$ (with a clipping of only 10 galaxies). 

\begin{figure}
\centering
\begin{minipage}{168mm}
\includegraphics[width=80mm]{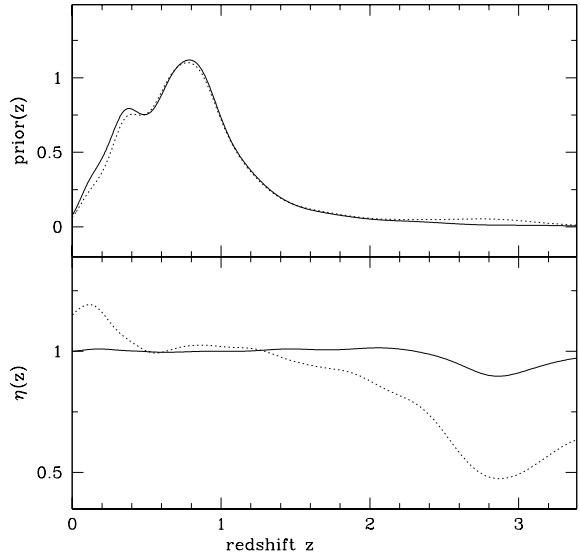}
\end{minipage}
\caption{The marginalized prior derived from the COSMOS sample of Scarlata et al. (2006), from which our spectroscopic sample is extracted, and its convergence properties. Upper panel: The marginalized prior after the first iteration (dotted line) and after five iterations (solid line) for a Gaussian smoothing length $\Gamma=0.05(1+z)$ in redshift space. Lower panel: The point-by-point ratio of two successive redshift-marginalized prior estimations. The dotted line shows the ratio of the prior estimates between the second and first iteration. The solid line shows the prior ratio between the fifth and fourth iteration.}
\label{fig:BayesConv}
\end{figure}

Similar results are found when comparing  the \zebra{} BY redshifts with the zCOSMOS spectroscopic redshifts.
In Figure~\ref{fig:BayesConv} we show the results of  the iterative calculation of the prior using the $>56000$ galaxies in the entire ACS-selected $I_{AB} \le 24$ COSMOS sample of \cite{scarlata2006} from which our spectroscopic sample was extracted. 
The prior was obtained using an adaptive Gaussian smoothing kernel of $\Gamma=0.05(1+z)$, which was tested to lead to a stable prior estimate.
In the figure, the left panel shows the prior estimate, marginalized to redshift space,  after one (dotted lines) and five (solid lines) iterations. Although we only present the prior marginalized over template types, the full 2D-prior is being used for the subsequent calculation of the posterior. The right panel shows the ratio $\eta(z)$ of the marginalized priors $P(z)$, from two successive iterations: the dotted lines correspond to ratio of the priors after the second and first iteration; the solid lines show the prior ratio between the fifth and fourth iterations.

\begin{figure}
\begin{minipage}{80mm}
\includegraphics[bb=15 140 564 690, width=80mm]{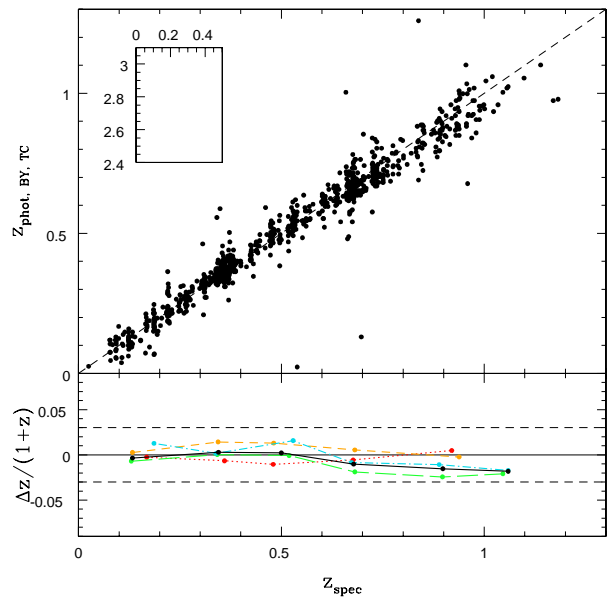}
\end{minipage}
\caption{The application of \zebra{}'s two-dimensional Bayesian method (i) after smoothing of the projected $P(z)$ prior with a smoothing scale of $\Gamma=0.05(1+z)$ and (ii) using the same template correction as in Fig.~\ref{fig:ResTempImp}. Symbols are as in Figure ~\ref{fig:ResTempNoImp} (except that no $\chi^2$ threshold is shown). Similar to the \zebra{} ML estimates {\it with template correction},  also the \zebra{} BY photo-z's {\it with template correction} eliminate the systematic trends that are present in the redshift estimates without template correction.}
\label{fig:ResTempImpBayes}
\end{figure}

In Figure~\ref{fig:ResTempImpBayes} we present the  \zebra-zCOSMOS  comparison as above,
this time  for the  \zebra{} BY redshift estimates {\it derived with template correction} ($z_{\text{phot,BY,TC}}$). These \zebra{} BY redshifts are obtained using  the values $z^\#$ and $t^\#$ which maximize the posterior. The figure highlights a similar high quality for the ML and BY \zebra{} estimates {\it with template correction}; indeed, the differences between the {\it template-corrected} BY and ML redshift estimates are vanishingly small. Of course, in BY mode \zebra{} returns the redshift and template probability distribution for each galaxy. The  BY run gives a 5-$\sigma$ clipped accuracy of $\sigma_{\Delta{}z/(1+z)}=0.027$ with only 7 outliers clipped, comparable to the one derived for the ML estimates.

\section{Concluding Remarks}
\label{sect:Conclusions}

A more thorough comparison of the \zebra{}  ML and BY photo-$z$'s with the currently available zCOSMOS redshifts is given by \cite{lil06}. Furthermore, the \zebra{} ML and BY photometric redshift estimates are compared by \cite{mobasher2006} with photo-$z$ estimates derived for the same galaxies with independent codes (these are either public codes, e.g. BPZ or have been developed by other teams within the COSMOS collaboration).

The \zebra{} ML photometric redshifts estimates for the COSMOS sample studied in this paper have already been used to derive the evolution up to $z\sim1$ of the luminosity functions for morphologically-classified early-, disk- and irregular-type galaxies (according to the classification scheme of {\it ZEST}, the Zurich Estimator of Structural Types; \citealt{scarlata2006}), to study the evolution up to similar redshifts of the number density of intermediate-size and large disk galaxies \citep{sargent2006} and to study the evolution of the luminosity function of elliptical galaxy progenitors \citep{scarlata2006a}.

The current version (1.0) of \zebra{} is being packaged with a user-friendly web-interface at the URL provided in the abstract. The \zebra{}  website will be constantly updated to provide the newest improved versions of the \zebra{} code, and the associated documentation describing in detail the implemented changes. In the meanwhile, \zebra{} is being upgraded with several new modules, including $(1)$ A module that incorporates dust absorption and reddening, according to several user-specifiable extinction corrections; $(2)$ An improved treatment of AGN's; and $(3)$ A module that uses several synthetic template models and a large choice of self-consistent star formation and metal enrichment schemes to estimate stellar masses, average ages and metallicities (and their uncertainties).

\section*{Acknowledgments}
We thank the anonymous referee for the helpful comments, which have improved the presentation of this paper and Manfred Kitzbichler for making available to the COSMOS collaboration the mock catalog used in this work.\\
This work is based on observations taken with: \\   - The NASA/ESA {\em
Hubble Space Telescope}, obtained at the Space Telescope Science
Institute, which is operated by AURA Inc, under NASA contract NAS
5-26555; \\ - The Subaru Telescope, which is operated by
the National Astronomical Observatory of Japan; \\ - Facilities at the European
Southern Observatory, Chile; \\ - Facilities at the Cerro Tololo Inter-American
Observatory and at the National Optical Astronomy Observatory, which are
operated by the Association of Universities for Research in Astronomy, Inc.
(AURA) under cooperative agreement with the National Science Foundation; and \\
- The Canada-France-Hawaii Telescope operated by the National Research Council of Canada, the Centre National de la Recherche Scientifique de France, and the University of
Hawaii.  \\  T. Lisker is acknowledged for helping with a 
preliminary reduction of a fraction of the groundbased near-infrared data.

\appendix

\section {Notation and definitions}
\label{sect:definitions}

The filter-averaged spectral flux density of a template can
be decomposed into a template-based spectral flux density
$f_{z,t,n}$ and a template normalization $a$ by setting:
\beq
af_{z,t,n}=
\frac{\int{}\frac{\dd\nu}{\nu}\thinspace{}s^{\text{obs}}_{\nu,t}(\nu)\thinspace{}n(\nu)}
{\int{}\frac{\dd\nu}{\nu}\thinspace{}n(\nu)}
\label{eq:PreDefFluxNu}
\eeq

The spectral flux density $s^{\text{obs}}_{\nu,t}$ measured in the observer frame is related to the rest-frame template
shape $s_{\nu,t}$  by:

\beq
s^{\text{obs}}_{\nu,t}(\nu/(1+z))=(1+z)\frac{L_\nu(\nu)}{4\pi{}D_L^2}=(1+z)s_{\nu,t}(\nu)a_t.
\label{eq:FluxTrans}
\eeq

The normalization factor $a_t$ matches the template shape $s_{\nu,t}(\nu)$ with the apparent spectral flux density
in the rest frame of a point-source with luminosity $L_\nu(\nu)$ at a luminosity distance $D_L$. The relation $a=a_t/(c(1+z))$ between $a$ and $a_t$ can then be derived from the above definitions. 

For high redshift sources, attenuation effects of intervening intergalactic material, especially neutral hydrogen, become increasingly important. These attenuation effects are mostly contributed by  Lyman series line-blanking and  photoelectric absorption\footnote{The component contributed by  photoelectric absorption is estimated by the approximation given in footnote 3 of \cite{1995ApJ...441...18M}.}.  These effects are accounted for by including a factor of $e^{-\tau(\lambda,z)}$ in the expression for $f_{z,t,n}$,  i.e., defining:
\beq
f_{z,t,n}=
\frac{\int{}\dd\lambda\lambda\thinspace{}s_{\lambda,t}(\lambda/(1+z))\thinspace{}e^{-\tau(\lambda,z)}\thinspace{}n(\lambda)}
{\int{}\dd\lambda/\lambda\thinspace{}n(\lambda)}.
\label{eq:DefFluxLambdaIGA}
\eeq
The \zebra{} user can choose to adopt either  the  \cite{1995ApJ...441...18M} or the \cite{2006MNRAS.365..807M} calculation for the attenuation term; compared with the former, the latter provides a somewhat lower absorption strength at any redshift. 
The K-correction $K_{nl}$ between filter band $l$ in the the rest-frame and filter band $n$ in the observed-frame is defined as:
\[
K_{nl}=m_n-M_l-5\log_{10}\left(D_L/10pc\right)
\]
with $M_l$  the absolute magnitude in the filter  $l$, 
and $m_n$  the apparent magnitude  in the filter band $n$.
The $K$-correction can be written as:
\begin{align*}
K_{nl}(z,t)&=\Mag(af_{z,t,n})-\Mag(f^{\text{em}}_{t,l}) \\
      &=\Mag(af_{z,t,n})-\Mag(a(1+z)f_{0,t,l}) \\
      &=2.5\log_{10}(1+z)+\Mag(f_{z,t,n})-\Mag(f_{0,t,l}).
\end{align*}
where $f^{\text{em}}_{t,n}$ is the restframe spectral flux density in the filter band $n$ and $\Mag(x)=-2.5\log_{10}(x)$. \zebra{} can provide the $K$-corrections for all (original, interpolated and corrected) used templates.

The \emph{normalized} likelihood $\mathcal{L}(z,t,a)$ can be written  as:
\begin{align*}
\mathcal{L}(z,t,a)
&=\frac{P(\mathbf{f}^\text{obs}|z,t,a)}{\sum_{t'}\int\dd{}z'\int{}\dd{}a'P(\mathbf{f}^\text{obs}|z',t',a')} \\
&=\frac{e^{-\frac{1}{2}\chi^2(z,t,a)}}{\sum_{t'}\int\dd{}z'\int{}\dd{}a'e^{-\frac{1}{2}\chi^2(z',t',a')}}
\end{align*}
with $P(\mathbf{f}|\boldsymbol{\alpha})$ as the conditional probability distribution of reproducing the data $\mathbf{f}$ given the parameters $\boldsymbol{\alpha}$.

The $\chi^2(z,t,a)$ can  be expressed as \citep{2000ApJ...536..571B}:
\beq
\chi^2(z,t,a)=F_{OO}-\frac{F_{OT}^2}{F_{TT}}+\left[a-\frac{F_{OT}}{F_{TT}}\right]^2F_{TT}
\label{eq:Chi2Benitez}
\eeq
where
\begin{align*}
F_{OO}&=\sum_{n=1}^{N_B}\left(\frac{f_n^\text{obs}}{\Delta_n}\right)^2,\\
F_{TT}&=\sum_{n=1}^{N_B}\left(\frac{f_{z,t,n}}{\Delta_n}\right)^2,\\
F_{OT}&=\sum_{n=1}^{N_B}\frac{f_n^\text{obs}\thinspace{}f_{z,t,n}}{(\Delta_n)^2}.
\end{align*}

In this  formulation, the best fitting template normalization $a^*$ is given by 
$a^*=F_{OT}/F_{TT}$. The best fitting redshift $z^*$ and template type $t^*$ follow from the maximum 
of $F_{OT}^2/F_{TT}$. The largest likelihood corresponds to the minimum
$\chi^2_{\text{min}}=F_{OO}-(F_{OT}^2/F_{TT})(z^*,t^*)$.

\section{The \zebra{}  $\chi^2$-minimization approach to template correction}
\label{sect:tempCorrection}

We first describe the simple case when only the  original set of templates is used as input, without interpolations between the original templates. We indicate with $N_t$ the number of catalog entries which are best fitted by a template type $t$.  In \cite{2000AJ....120.1588B} the spectral distribution $s_t^{\text{orig}}(k)$ of the original template type $t$ is changed by a $\chi^2$-minimization over all template shapes $s_t^{\text{cor}}(k)$, iteratively for all entries $i\in{}{N}_t$. Specifically,  \cite{2000AJ....120.1588B}  perform the template correction by minimizing the following $\chi^2$ function:
\[
\chi_{t,i}^2=\sum_k\frac{1}{\sigma_{t,k}^2}(s_t^{\text{cor}}(k)-s_t^{\text{orig}}(k))^2+
\sum_{n=1}^{N_B}\frac{1}{\Delta_{n,i}^2}(f_{n,i}^{\text{cor}}-f_{n,i}^{\text{obs}})^2
\]

In our approach, the shape $s_t^{\text{orig}}(t)$ of a given basic template $t$ is changed in one step, taking all entries $i\in{}{N}_t$ into account at once; furthermore, a regularization term is included in the definition of $\chi^2$, to avoid unphysical high frequency fluctuation in the correction of the template as a function of wavelength. We therefore determine the optimal template corrections by minimizing:

\begin{align}
\label{eq:chi2Minim}
\chi_t^2&=\frac{1}{N_t}\sum_{i=1}^{N_t}\chi_{t,i}^2=\sum_k\frac{1}{\sigma_{t,k}^2}(s_t^{\text{cor}}(k)-s_t^{\text{orig}}(k))^2 \nonumber \\
&+\frac{1}{N_t}\sum_{i=1}^{N_t}\sum_{n=1}^{N_B}\frac{1}{\Delta_{n,i}^2}(f_{n,i}^{\text{cor}}-f_{n,i}^{\text{obs}})^2 \nonumber \\
&+\sum_k\frac{1}{\rho_{t,k}^2}(s_t^{\text{cor}}(k+1)-s_t^{\text{cor}}(k)-s_t^{\text{orig}}(k+1)+s_t^{\text{orig}}(k))^2 
\end{align}

with the variables as described in Section \ref{sect:tempCorr}.

The spectral flux density $f_{n,i}^{\text{cor}}$ of the corrected template in the 
filter band $n$ depends on the catalog entry $i$ through its best fitted template type $t$, redshift $z$ and normalization factor $a$. Specifically:
\beq
\label{eq:chi2Flux}
f_{n,i}^{\text{cor}}=\sum_k T_n^i(k)s_{t}^{\text{cor}}(k),
\eeq
where $T_n^i(k)s_{t}^{\text{cor}}(k)$ has yet to be determined.

In the Maximum Likelihood procedure (Section \ref{sect:ML}), the template-based spectral flux density $f_{t,z,n}$ is calculated  for each template $t$, filter $n$ and redshift $z$, modulo an  overall normalization constant $a$. The procedure assigns to each entry $i$ a triple $(t(i),z(i),a(i))$ so that  the $\chi^2$ is minimized.

\zebra{} uses a {\it linear approximation} to describe  the spectral flux density through the best fit template shape, i.e.:
\begin{align*}
f_{n,i}&=a(i)f_{z(i),t(i),n}\approx{}\sum_k T_n^i(k)s_{t(i)}(k),\text{ with} \\
T_n^i(k)&=\frac{(1+z(i))^2\thinspace{}a(i)}{\int{}\dd\lambda{}/\lambda{}\thinspace{}n(\lambda)}\Delta\lambda\lambda_k\thinspace{}n(\lambda_k(1+z(i)))
\end{align*}

The effect of intergalactic absorption is  included easily by extending the definition of 
$T_n^i(k)$ using \eqref{eq:DefFluxLambdaIGA}:
\[
T_n^i(k)=\frac{(1+z(i))^2\thinspace{}a(i)}{\int{}\dd\lambda{}/
\lambda{}\thinspace{}n(\lambda)}\Delta\lambda\lambda_k\thinspace{}
n(\lambda_k(1+z(i)))\thinspace{}e^{-\tau(\lambda_k(1+z(i)),z(i))}
\]

The two-step iterative template correction then proceeds as  described in Section~\ref{sect:tempCorr}.

When log-interpolated templates are used we define the set $\mathcal{N}_t$ as the set of catalog entries $i$, so that  $t$ is the nearest basic type of the  best fitting type $t(i)$.  If $t(i)$ is a basic template,  then $t=t(i)$; if $t(i)$ is an interpolated template, then  $t=t_1$ if $g<0.5$, or otherwise $t=t_2$, see 
\eqref{eq:defLogInterpol}. To simplify the notation we  define:
\[
s^{\text{log}}_{t,g}=\begin{cases}
	s^{\text{log}}_{t,t^+,g} & \text{ if }g<0.5 \\
	s^{\text{log}}_{t^-,t,g} & \text{ if }g\geq{}0.5\text{.}
	\end{cases}
\]
Here $t^+$ and $t^-$ indicate the successor and predecessor basic template of the basic template type $t$ with respect to the (assumed)
global ordering. 

When using interpolated templates, equation \eqref{eq:chi2Flux} has to be re-defined. 
In particular, a change in the shape $s^{\text{orig}}_{t}(k)$ leading to 
$s^{\text{cor}}_{t}(k)$ is reflected in a changed spectral flux density $f_{n,i}^{\text{cor}}$ for each entry $i\in{}\mathcal{N}_t$.   For  $g(i)<0.5$, we obtain:

\begin{align*}
f_{n,i}^{\text{cor}}
&=\sum_k T_n^i(k)s^{\text{log}}_{t(i),g(i)} \\
&=\sum_k T_n^i(k)(s^{\text{cor}}_{t}(k))^{1-g(i)}(s_{t^+}(k))^{g(i)}
\end{align*}

Using:

\[
(s^{\text{cor}}_{t}(k))^{1-g(i)}=(s^{\text{orig}}_{t}(k))^{1-g(i)}\left(1+\frac{\xi_t(k)}{s^{\text{orig}}_{t}(k)}\right)^{1-g(i)}
\]

and assuming that $\xi_t(k)=s^{\text{cor}}_{t}(k)-s^{\text{orig}}_{t}(k)$ is  small in comparison with $s^{\text{orig}}_{t}(k)$, the following approximation holds:

\beq
\label{eq:fluxglt05}
f_{n,i}^{\text{cor}}\approx{}
\sum_k T_n^i(k)(s_{t^+}(k))^{g(i)}(s^{\text{orig}}_{t}(k))^{1-g(i)}\left(1+(1-g(i))\frac{\xi_t(k)}{s^{\text{orig}}_{t}(k)}\right)
\eeq

Similarly, for $g(i)\geq{}0.5$, we obtain:

\beq
\label{eq:fluxggt05}
f_{n,i}^{\text{cor}}\approx{}
\sum_k T_n^i(k)(s_{t^-}(k))^{1-g(i)}(s^{\text{orig}}_{t}(k))^{g(i)}\left(1+g(i)\frac{\xi_t(k)}{s^{\text{orig}}_{t}(k)}\right)
\eeq

In this approximation the spectral flux density depends \emph{linearly} on $\xi_t(k)$ and $s^{\text{cor}}_{t}(k)$, respectively.
With the definition $s^{\text{log}}_{t,0}=s^{\text{log}}_{t,1}=s^{\text{cor}}_{t}(k)$, the equations 
\eqref{eq:fluxglt05} and \eqref{eq:fluxggt05} also describe the change in spectral flux density if the best fit template is an original template.

To minimize the $\chi^2$, the templates are sampled  on a grid 
linearly spaced in units of $\log(\lambda)$; all templates are  normalized to the spectral flux density of unity in the B-band, in order to be able to use for each template the same pliantness $\sigma$.
With the definitions:

\begin{align*}
g_n^i&=
\begin{cases}
f_{n,i}^{\text{obs}}-\sum_k T_n^i(k)(s_{t^+}(k))^{g(i)}(s^{\text{orig}}_{t}(k))^{1-g(i)}&\text{ if }g(i)<0.5 \\
f_{n,i}^{\text{obs}}-\sum_k T_n^i(k)(s_{t^-}(k))^{1-g(i)}(s^{\text{orig}}_{t}(k))^{g(i)}&\text{ if }g(i)\geq{}0.5
\end{cases} \\
c_n^i(k)&=
\begin{cases}
T_n^i(k)(s_{t^+}(k))^{g(i)}(s^{\text{orig}}_{t}(k))^{-g(i)}(1-g(i))&\text{ if }g(i)<0.5 \\
T_n^i(k)(s_{t^-}(k))^{1-g(i)}(s^{\text{orig}}_{t}(k))^{g(i)-1}g(i)&\text{ if }g(i)\geq{}0.5
\end{cases}
\end{align*}

equation \eqref{eq:chi2Minim} can be written as:

\begin{align}
\label{eq:chi2MinimNew}
\chi_t^2&=\sum_k\frac{1}{\sigma_{t,k}^2}(\xi_t(k))^2+\sum_k\frac{1}{\rho_{t,k}^2}(\xi_t(k+1)-\xi_t(k))^2 \nonumber \\
&\phantom{=}+\frac{1}{N_t}\sum_{i=1}^{N_t}
\sum_{n=1}^{N_B}\frac{1}{\Delta_{n,i}^2}(g_n^i-\sum_k c_n^i(k)\thinspace{}\xi_t(k))^2
\end{align}

Postulating
$\frac{\partial}{\partial{}\xi_t(l)}\chi_t^2=0$ leads to a system of linear equations in $\xi_t(k)$, i.e.:

\[
	\sum_kM_t(l,k)\thinspace{}\xi_t(k)=\nu_t(l),
\]

where

\begin{align}
M_t(l,k)&=\frac{\delta_{l,k}}{\sigma_{t,k}}+\frac{1}{N_t}\sum_{i=1}^{N_t}\sum_{n=1}^{N_B}
\frac{1}{\Delta_{n,i}^2}(c_n^i(k)\thinspace{}c_n^i(l)) \nonumber \\
&\phantom{=}+\frac{1}{\rho_{t,k-1}}(\delta_{l,k}-\delta_{l,k-1})+\frac{1}{\rho_{t,k}}(\delta_{l,k}-\delta_{l,k+1})\\
\nu_t(l)&=\frac{1}{N_t}\sum_{i=1}^{N_t}\sum_{n=1}^{N_B}\frac{1}{\Delta_{n,i}^2}(c_n^i(l)\thinspace{}g_n^i)
\end{align}

The density of the $\lambda$-grid used to sample the templates  determines the size of the set of linear equations.  In the application to the COSMOS sample described in Section~\ref{sect:application},  we have used a grid in $\log(\lambda)$-space  of about 800 points.

Attention  has to be paid in carefully choosing the free parameters, in order to obtain physically meaningful corrections to the templates when using also  interpolated templates. Specifically, if the absolute change  $|\xi_t(k)|$  of a template is larger than the value $s^{\text{orig}}_{t}(k)$ of the original  template at that wavelength, the approximative treatment of the log-interpolated templates becomes inappropriate. This can happen if  a too high pliantness $\sigma_{t,k}$ is used, and/or if too few galaxies are available to constrain the fits that are performed to correct the templates.  If a corrections would make the flux of a template negative in some wavelength region, the flux is set to zero. If that happens,  the $c_n^i(k)$ coefficient is also set to zero, thereby inhibiting any further change in that template at that specific wavelength.

\section{Testing \lzebra{} on a mock sample}
\label{sect:mock}
We further  demonstrate the performance of \zebra{} using a mock catalog that has been produced for the COSMOS field. Simulations of galaxies rely on population synthesis and dust models which may not perfectly match the observed SEDs of real galaxies. We find indeed that the use of the galaxy templates discussed in Section \ref{sect:dataTemplates} provides slightly less accurate photometric redshift estimates  for the mock galaxies than for real data. On the other hand, adopting the same models that were used to construct  the mock galaxies when recovering their photo-$z$'s results in unrealistically accurate results. Testing the code on a mock catalog has however several advantages, as the mock catalog provides a large set of data with known precise redshifts, and hence allows us to test  the reliability and stability of the code using disjoint samples for the training set (that is used for the template correction) and for the assessment of the photo-$z$ accuracy.

The mock catalog used for our tests contains about 50000 galaxies with $I\le22.5$ and data in five photometric bands ($B$, $g$, $i$, $r$, $K_s$).  We used the same templates discussed in Section \ref{sect:dataTemplates}. In order to directly compare the results obtained with the mock data  with those obtained using the zCOSMOS spectroscopic redshifts, we limited the training set to  1000 mock galaxies, and we used a sample of 10000 mock galaxies, disjoint from the training set, to perform the tests. 

A run of \zebra{} in \pcm{} on the original mock data showed that systematic photometric offsets were smaller than the assumed relative photometric error of 0.05 magnitudes. To test the effect of  systematic photometric offsets,  we therefore added shifts up to 0.2 mag to the mock data. These offsets were  correctly identified and removed by the \zebra{}'s \pcm{}. 

 \zebra{} was then run in the \tom{}; this was done using, for the training set,   photometric data both corrected and not corrected for the added "extra" offsets discussed above, so to establish the impact of systematic photometric errors on the template correction procedure.  The entire set of original plus corrected templates was then used in the analysis. 
  
In Table \ref{tab:ResMock} we summarize the results of applying the \mlm{} of \zebra{} both to recover the redshifts of the training set galaxies themselves, and to estimate the redshifts of the independent set of 10000 galaxies in the "evaluation" catalog.

\begin{figure}
\begin{minipage}{168mm}
\includegraphics[width=80mm]{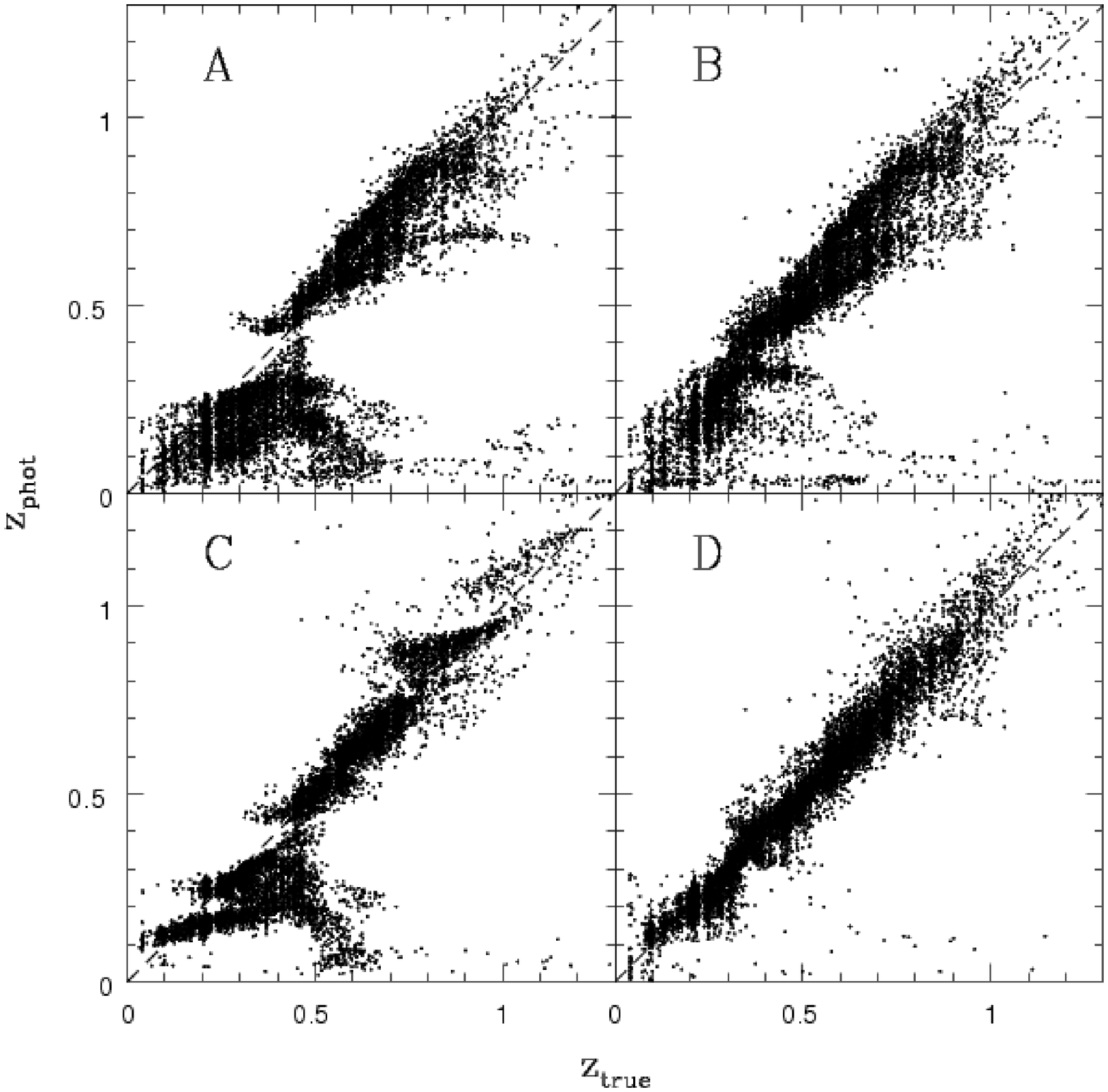}
\end{minipage}
\caption{The comparison between the \zebra{} photometric redshifts and  the ``true'' redshifts of  the mock galaxies. The figure refers to the ``evaluation'' runs in which 1000 galaxies are used as the training set, and the evaluation of the performance is made on a non-overlapping sample of 10000 mock galaxies. Four different cases are shown:  (a)  The catalogs contains substantial photometric offsets, and the templates are not optimized; (b) The photometry correction scheme is now applied, but no template optimization has yet been performed; (c) No photometric correction is performed,  but the template optimization scheme has been applied; (d) Photometric errors are removed from both the evaluation sample and the training sample, and  the template optimization scheme is applied.}
\label{fig:mockComp}
\end{figure}

Four configurations were explored, i.e., using: $(a)$ Catalog not corrected for photometric offsets and original (``uncorrected'') templates; $(b)$ Catalog corrected for systematic photometric errors and again original, uncorrected templates; $(c)$ Catalog not corrected for photometric offsets and corrected/optimized templates; (d) Catalog corrected for systematic photometric errors and corrected/optimized templates. In Figure~\ref{fig:mockComp} we compare the resulting photometric redshifts for the 10000 galaxy  ``evaluation sample''. 

\begin{table}
\begin{tabular}{|c|c|c|c|c|c|}
\hline
Catalog & Phot.  &  Templ. & $\sigma$ & $\Delta{}z/(1+z)$ & $\%$ \\
 & corr.  &  optim. & & &  \\\hline \\
Training & no & no & 0.1008 & -0.051 & 1.3 \\
Training & yes & no & 0.0526 & -0.001 & 2.7 \\
Training & no & yes &  0.0785 & -0.029 & 0.7 \\
Training & yes & yes & 0.0345 & 0.000 & 1.0 \\
Evaluation & no & no &  0.1004 & -0.050 & 1.2 \\
Evaluation & yes & no &  0.0590 & -0.001 & 2.4\\
Evaluation & no & yes &  0.0780 & -0.029 & 0.5 \\
Evaluation & yes & yes & 0.0350 & -0.001 & 1.4 \\
\end{tabular}
\caption{Results of the application of  \zebra{} in \mlm{}  to 
1000 mock galaxies that are also used as training set (``Training'' catalog), and of the application of the code to a sample of
10000 mock galaxies (``Evaluation'' catalog) not overlapping with the ``Training'' catalog. The second column indicates whether the photometric catalogs are corrected for systematic errors;  the third column indicates whether  the template-correction scheme has been applied. Columns four and five list the accuracy $\sigma_{\Delta{}z/(1+z)}$  and the mean offset $\Delta{}z/(1+z)$ of the photometric redshift when compared with the ``true'' redshifts after 5-$\sigma$ clipping. The percentage of 5-$\sigma$ outliers is listed in the last column. Note the high accuracy and lack of global shift that is obtained when both the corrections to the photometric catalogs and the template optimization are applied; also, accuracies of the same order are obtained in the ``Training'' and ``Evaluation''  runs.}
\label{tab:ResMock}
\end{table}

These tests indicate that:
\begin{enumerate}
\item The accuracies  of the photometric redshifts obtained when applying  \zebra{} to the galaxies of the training sample itself and to the disjoint evaluation sample are nearly identical (see Table \ref{tab:ResMock}). This shows that results of the \pcm{} and \tom{} are robust and lead to a high accuracy in the redshift estimates;
\item Systematic photometric errors may indeed lead to substantial systematic artefacts in the photometric redshift estimates, which need to be removed before the template correction is performed;
\item Accurate redshifts without significant systematic artefacts can only be achieved if both photometric corrections and template corrections are employed.
\end{enumerate}

\bsp

\label{lastpage}

\end{document}